\documentclass[11pt,a4paper]{article}

\usepackage{amsmath, amssymb, amsthm, graphicx, epsfig, fancyhdr,epsfig, slashed, mathrsfs}
\usepackage{subfigure}
\usepackage{tikzsymbols}
\usepackage{float}
\usepackage{jheppub}

\usepackage{mathtools} 

\usepackage{tikz,xcolor,hyperref}
\usepackage{siunitx}
\DeclareSIUnit \parsec {pc}
\usepackage{braket}
\usepackage{csquotes}

\usepackage{amsfonts}
\usepackage{mathtools}
\usepackage{pifont}

\newcommand{\benum}{\begin{enumerate}}
\newcommand{\eenum}{\end{enumerate}}
\newcommand{\bi}{\begin{itemize}}
\newcommand{\ei}{\end{itemize}}

\newcommand{\be}{\begin{equation}}
\newcommand{\ee}{\end{equation}}

\newcommand{\bea}{\begin{eqnarray}}
\newcommand{\eea}{\end{eqnarray}}
\newcommand{\beq}{\begin{eqnarray}}
\newcommand{\eeq}{\end{eqnarray}}

\newcommand{\Rmnum}[1]{\expandafter\@slowromancap\romannumeral #1@}

\definecolor{lime}{HTML}{A6CE39}
\DeclareRobustCommand{\orcidicon}{
	\begin{tikzpicture}
	\draw[lime, fill=lime] (0,0) 
	circle [radius=0.2] 
	node[white] {{\fontfamily{qag}\selectfont \tiny ID}};
	\draw[white, fill=white] (-0.0625,0.095) 
	circle [radius=0.007];
	\end{tikzpicture}
	\hspace{-2mm}
}

\foreach \x in {A, ..., Z}{\expandafter\xdef\csname orcid\x\endcsname{\noexpand\href{https://orcid.org/\csname orcidauthor\x\endcsname}
			{\noexpand\orcidicon}}
}

\allowdisplaybreaks

\title{
Impact of high-scale Seesaw and Leptogenesis on inflationary tensor perturbations as detectable   gravitational waves }

\author[a]{Maximilian Berbig,}
\author[b]{Anish Ghoshal}
\affiliation[a]{Bethe Center for Theoretical Physics und Physikalisches Institut der Universitt Bonn,\\Nussallee 12, Bonn, Germany}
\affiliation[b]{Institute of Theoretical Physics, Faculty of Physics, University of Warsaw,\\
ul. Pasteura 5, 02-093 Warsaw, Poland}
\emailAdd{berbig@physik.uni-bonn.de}
\emailAdd{anish.ghoshal@fuw.edu.pl}

\abstract{We discuss the damping of inflationary gravitational waves (GW) that re-enter the horizon before or during an epoch, where the energy budget of the universe is dominated by an unstable right handed neutrino (RHN), whose out of equilibrium decay releases entropy. Starting from the  minimal Standard Model extension, motivated by the observed neutrino mass scale, with nothing more than 3 RHN for the Seesaw mechanism, we discuss the conditions for high scale leptogenesis assuming a thermal initial population of RHN. We further address the associated production of potentially light non-thermal dark matter and a potential component of dark radiation from the same RHN decay. One of our main findings is that the frequency, above which the damping of the tensor modes is potentially observable, is completely determined by successful leptogenesis and a Davidson-Ibarra type bound to be at around $\SI{0.1}{\hertz}$. To quantify the detection prospects of this GW background for various   proposed  interferometers such as \textsc{AEDGE}, \textsc{BBO}, \textsc{DECIGO}, \textsc{Einstein Telescope} or \textsc{LISA} we compute the \textit{signal-to-noise ratio} (SNR). This allows us to investigate  the viable parameter space of our model, spanned by the mass of the decaying RHN $M_1 \gtrsim \SI{2.4e8}{\giga\electronvolt} \cdot \sqrt{\SI{2e-7}{\electronvolt}/\tilde{m}_1}$ (for leptogenesis) and the effective neutrino mass parameterizing its decay width $\tilde{m}_1< \SI{2.9e-7}{\electronvolt}$ (for RHN matter domination).  Thus gravitational wave astronomy is a novel way to probe both the  Seesaw and the leptogenesis scale, which are  completely inaccessible to laboratory experiments in high scale scenarios.}

 
\begin{document}
\maketitle

\section{Introduction}
\label{intro}
 The standard model (SM) of particle physics predicts that the neutrinos are massless, but due to the observation of neutrino oscillations for solar \cite{dc27cfb,Super-Kamiokande:2001bfk,Super-Kamiokande:2002ujc,SNO:2002tuh,Super-Kamiokande:2005mbp,articleKam, PhysRevD.94.052010,Borexino:2015axw}, atmospheric \cite{IceCube:2017lak, ANTARES:2018rtf} and reactor \cite{KamLAND:2008dgz,T2K:2011ypd,DoubleChooz:2011ymz,T2K:2013ppw} neutrinos we now know that they are massive and the flavor states mix due to the  propagation of multiple mass eigenstates. Moreover 
 the $\beta$-decay experiment KATRIN \cite{KATRIN:2021uub} has provided us with the first direct limit of the neutrino mass scale $m_\nu < \SI{0.8}{\electronvolt}$. 
 Cosmology offers an indirect probe of this scale and demands that the sum of all neutrino masses satisfies $\sum_i m_{\nu_i} < \SI{0.12}{\electronvolt}$ \cite{Aghanim:2018eyx,eBOSS:2020yzd} in order to be consistent with the predictions for the Cosmic Microwave Background (CMB) radiation, Large scale structure (LSS) formation and  Big Bang Nucleosynthesis (BBN).  The accelerated expansion at the beginning of the universe provided by cosmic inflation, which was postulated in order to solve the horizon and the flatness problems and is responsible for quantum generation of the primordial fluctuations seeding the large scale structure   of the universe, is thought to be driven by a scalar field known as the inflaton (see \cite{Martin:2013tda} for a review). In this paper, we will be concerned with the primordial Gravitational Waves (GW) background of such inflationary origin \cite{Grishchuk:1974ny,Starobinsky:1979ty,Rubakov:1982df} (see \cite{Guzzetti:2016mkm} for a review on this topic).
These inflationary GWs can act as a logbook of the expansion history of our universe throughout its entire
evolution~\cite{Seto:2003kc,Boyle:2005se,Boyle:2007zx,Kuroyanagi:2008ye,Nakayama:2009ce,Kuroyanagi:2013ns,Jinno:2013xqa,Saikawa:2018rcs}.
Particularly, the detailed time evolution of the Hubble rate during
the expansion determines the transfer function that describes how
gravitational waves at different frequencies are red-shifted to the present day.
This property turns primordial GWs into a powerful tool that grants access to
the thermal history of our universe prior to BBN. Primordial GWs offer, \textit{e.g.} an opportunity to measure  the reheating temperature after inflation~\cite{Bernal:2020ywq,Nakayama:2008ip,Nakayama:2008wy,
Kuroyanagi:2011fy,Buchmuller:2013lra,Buchmuller:2013dja,Jinno:2014qka,Kuroyanagi:2014qza}.
Similarly, with help of these inference can be drwan of the equation of state during
the quark-hadron phase transition in quantum
chromodynamics~\cite{Schettler:2010dp,Hajkarim:2019csy} or
constrain properties of the hidden sectors beyond the Standard Model (BSM) of particle physics~\cite{Jinno:2012xb,Caldwell:2018giq}.\\
\\
The observed baryon asymmetry of the universe (BAU) is longstanding puzzle in particle physics and cosmology  \cite{Zyla:2020zbs,Aghanim:2018eyx}. While the universe is expected to start in a matter-antimatter symmetric phase, any primordial asymmetry set due tothe initial conditions is expected to get diluted by the exponential expansion phase during cosmic inflation. The BAU is often quoted in terms of the  baryon to photon ratio measurement which, according to the latest Planck 2018 data, is given by \cite{Aghanim:2018eyx} 
\begin{equation}
\eta_B = \frac{n_{B}-n_{\bar{B}}}{n_{\gamma}} = 6.1 \times 10^{-10}
\label{etaBobs}
\end{equation}
and agrees with the value extracted from BBN \cite{Fields:2019pfx} as well.
Similar to the BAU, there has been another question related to the presence of a mysterious, non-luminous form of matter, popularly known as dark matter (DM), giving rise to approximately $26\%$ of the energy density in the present universe. In terms of density 
parameter  $\Omega_{\rm DM}$ and $h = H_0/(100 \;\text{km} ~\text{s}^{-1} 
\text{Mpc}^{-1})$ with $H_0$ being the observed present day  Hubble parameter, the current DM abundance is conventionally reported to be \cite{Aghanim:2018eyx}
\begin{equation}
\Omega_{\text{DM}} h^2 = 0.120\pm 0.001
\label{dm_relic}
\end{equation}
at 68\% CL. Apart from cosmological evidence, the presence of DM has also been suggested by several astrophysical implications \cite{Zwicky:1933gu, Rubin:1970zza, Clowe:2006eq}. While none of the standard model particles satisfy the criteria of a particle DM candidate, the SM also does not to satisfy the criteria  to dynamically generate the observed BAU, known as Sakharov's conditions \cite{Sakharov:1967dj}, in adequate amounts. This has led to several BSM possibilities offering intriguing solutions to these puzzles: The Type I Seesaw mechanism \cite{Minkowski:1977sc,Yanagida:1979as,Gell-Mann:1979vob,Glashow:1979nm,10.1143/PTP.64.1103, PhysRevLett.44.912}, where the SM is augmented with three right handed SM gauge singlet neutrinos (RHN),  may explain both the observed neutrino masses (from neutrino oscillation experiments) as well as the baryon asymmetry of the universe via first generating an asymmetry in the dark leptonic sector \cite{Fukugita:1986hr,Luty:1992un,Plumacher:1996kc,Covi:1996wh,Giudice:2003jh} and subsequently getting transferred to the visible baryonic sector via the electroweak sphaleron transitions \cite{Kuzmin:1985mm}.  Among the BSM proposals for DM, the weakly interacting massive particle (WIMP) \cite{Kolb:1990vq} produced as a thermal relic is perhaps the most widely studied one (see  \cite{Arcadi:2017kky} for a review). However due to the absence of any WIMP related signals in nuclear and electron recoil DM direct detection experiments, there has been growing interest in other (non-thermal) production modes: some examples are the well-known super-WIMP scenario \cite{Feng:2003xh}, where frozen out WIMP decays to the actual DM, FIMPs \cite{Hall:2009bx} (see \cite{Bernal:2017kxu} for a review) that have such tiny couplings to the SM plasma that they never thermalize, or non-thermal production  from inflaton decays \cite{Gelmini:2006pw} during the process of the formation of the radiation bath known as reheating. In leptogenesis models the  RHN might also have the decay modes to other SM singlets that can be good DM candidates \cite{Falkowski:2011xh,Falkowski:2017uya}, which is why we will adopt this framework. Since the RHN decays out-of-thermal equilibrium the DM will be non-thermal.
\\
\\




\noindent
We will demonstrate that the same RHN decay responsible for both the generation of the primordial baryon asymmetry via leptogenesis, as well as the production of non-thermal dark matter  and a   possible component of dark radiation, leaves its vestige on the primordial spectrum of inflationary GWs. In particular we consider an epoch of intermediate matter domination  \cite{PhysRevD.31.681,Kolb:1990vq,Bezrukov:2009th}  from the lightest RHN, which decouples from the plasma while relativistic and is very long-lived compared to the characteristic time scale of the cosmic expansion.  Since the  decay occurs far away from thermal equilibrium it will release a large amount of entropy, which dilutes the energy density of primordial GWs that enter the horizon before the decay.\\
\\
\noindent  Although the Seesaw mechanism ties leptogenesis to the observed light neutrino masses, the mechanism itself is notoriously difficult to test in laboratory based experiments, as the heavy right-handed neutrino mass scale has to be above $ \gtrsim  10^{9}$~GeV (see \cite{Buchmuller:2004nz}). One should keep in mind that this bound can be evaded, see for example \cite{Pilaftsis:2003gt} and with some fine tuning, it is also possible to bring down the scale of the non-resonant thermal leptogenesis to as low as $10^6$ GeV \cite{Moffat:2018wke}. However indirect tests for high scale leptogenesis of course exist as well. These are primarily based on neutrino-less double beta decay scenarios \cite{Schechter:1981bd,DellOro:2016tmg}, meson decay scenarios \cite{Shrock:1980vy,Kayser:1981nw,DeVries:2020jbs}, and via CP violation in the neutrino oscillation \cite{Endoh:2002wm,Esteban:2016qun},  the structure of the leptonic mixing matrix \cite{Bertuzzo:2010et}, or via considering theoretical constraints from the demand of the SM Higgs vacuum does not become unstable in early universe \cite{Ipek:2018sai,Croon:2019dfw}.
Therefore, it is necessary, although very challenging to find newer and complementary tests of such heavy neutrino seesaw physics and consequently the leptogenesis mechanism. Recently it has been proposed to complement these indirect tests with the observations of GWs of primordial origin such as that from cosmic strings \cite{Dror:2019syi},  domain walls \cite{Barman:2022yos} and other topological defects \cite{Dunsky:2021tih} or from nucleating and colliding vacuum bubbles \cite{Dasgupta:2022isg,Borah:2022cdx}, graviton bremmstrahlung \cite{Ghoshal:2022kqp} and primordial black holes \cite{Bhaumik:2022pil,Bhaumik:2022zdd}.
These previous studies on GW \cite{PhysRevD.31.3052,Buchmuller:2013lra,Chao:2017ilw,Okada:2018xdh,Buchmuller:2019gfy,Hasegawa:2019amx,Haba:2019qol,Dror:2019syi,Blasi:2020wpy,Dunsky:2021tih} focused on the stochastic GW background from the  dynamics of the scalar field, whose vacuum expectation value is responsible for the RHN mass, whereas (when it comes to leptogenesis) we only extend the  SM by adding nothing more than three RHNs with hard mass terms. In order to ensure a thermal population of the lightest RHN, which can not be established by the  Yukawa couplings we consider, we have to assume that the RHNs are produced from inflaton decays or additional gauge interactions. In this paper we propose the imprint of the RHN decay on the  inflationary first-order tensor perturbations as a novel probe of the minimal high-scale leptogenesis mechanism. \\
\\
\noindent \textit{The paper is organized as follows:} In the subsection  \ref{sec:basic} of section \ref{sec:2} we discuss   the Seesaw model, then how the decay of the lightest right handed neutrino (RHN) leads to an intermediate era of matter domination in \ref{sec:long},  and we elaborate on  the generation of baryon asymmetry via leptogenesis from the decay of the lightest RHN in \ref{sec:lepto}. We also discuss the production of non-thermal dark matter and dark radiation from such heavy RHN decays in \ref{sec:DM}. In section \ref{sec:3} we discuss the generation and propagation of inflationary tensor perturbations as Gravitational Wave signals and show how RHN decays leave their imprint on the GW spectrum. We discuss the GW detection prospects in \ref{sec:res} of  section \ref{sec:4} 
and translate such experimental sensitivities into the reach for probing the parameter space and scale of leptogenesis via computing the  \textit{signal-to-noise ratio} (SNR) in \ref{sec:SNR}. We  end with the  conclusions   in section \ref{sec:5}.


\medskip

\section{Decays of a long-lived RHN}\label{sec:2}
\subsection{Type I Seesaw mechanism}\label{sec:basic}
\noindent We start with  a conventional Type I Seesaw  \cite{Minkowski:1977sc,Yanagida:1979as,Gell-Mann:1979vob,Glashow:1979nm,10.1143/PTP.64.1103, PhysRevLett.44.912} with three right handed neutrinos $N$
\begin{align}\label{eq:Seesaw}
    \mathcal{L} = \lambda\; \overline{L} (i\sigma_2) H^\dagger N + \frac{M_N}{2}\; \overline{N^c} N +\text{h.c.},
\end{align}
where $\sigma_2$ is the second Pauli matrix and assume without loss of generality that the symmetric right handed neutrino (RHN) mass matrix is diagonal
\begin{align}
    M_N = \text{diag}\left(M_1,M_2,M_3\right),
\end{align}
without making any assumptions about the mass spectrum yet. After Integrating out the RHN and electroweak symmetry breaking with $\braket{H}\equiv v= \SI{174}{\giga\electronvolt}$  the active neutrino mass matrix reads at leading order in the Seesaw expansion 
\begin{align}
    m_\nu = - m_D \cdot M_N^{-1} \cdot m_D^t= \text{diag}\left(m_1,m_2,m_3\right),  \quad \text{with} \quad m_D \equiv  \lambda \;v \ll M_N.\label{eq:seesaw}
\end{align}
Using the Casas-Ibarra parameterization in the basis where the charged lepton mass matrix is diagonal one finds \cite{Casas:2001sr}
\begin{align}
    \lambda = \frac{1}{v}\cdot M_N^\frac{1}{2}\cdot R \cdot m_\nu^\frac{1}{2}\cdot U_\text{PMNS}^\dagger,
\end{align}
where $U_\text{PMNS}$ is the leptonic equivalent of the CKM matrix. $R$ describes the mixing and  CP-violation in the RHN sector and is expressed as a complex, orthogonal matrix that reads
\begin{align}
    R \equiv \text{diag}(\pm 1,\pm 1,\pm 1)\cdot  R^{(23)}(z_{23})\cdot  R^{(13)}(z_{13})\cdot  R^{(12)}(z_{12}) 
\end{align}
in terms of $2\times 2$ rotation matrices $R^{(ij)}$ in the $ij$-plane with an angle $z_{ij}$.

\subsection{Conditions for intermediate matter domination}\label{sec:long}
\noindent The lightest RHN $N_1$ has the tree level decay width summed over all SM lepton flavours of
\begin{align}\label{eq:dec}
    \Gamma_1\equiv \Gamma(N_1\rightarrow L H, \overline{L} H^\dagger) = \frac{\left|\lambda \lambda^\dagger\right|_{11}}{8\pi} M_1. 
\end{align}
For $T\gg M_j$ the decay in the plasma is suppressed by a time dilation factor of $M_1/T$ \cite{Kolb:1979qa}, which goes to one for $T\leq M_1$.
It is customary to define the effective neutrino mass mediated by $N_1$ 
\begin{align}
    \tilde{m}_1 \equiv \frac{\left|\lambda^\dagger \lambda\right|_{11} v^2}{M_1}= \sum_{i} m_i |R_{1i}|^2,
\end{align}
which appears when  comparing the decay rate to  the characteristic time scale of cosmic expansion $H(T)^{-1}$, where $H(T)$ is the Hubble rate during radiation domination
\begin{align}
    K_1\equiv \frac{\Gamma_1}{2 H(T)}\Big|_{T=M_1} = \frac{\tilde{m}_1 }{\SI{2e-3}{\electronvolt}}.
\end{align}
This effective mass only coincides with the physical mass ($\tilde{m}_j=m_j$) for $R_{ji}=0,\forall i\neq j$.
A small effective mass $\tilde{m}_1$ implies that $N_1$ is weakly coupled to other two RHN. 
One can show that this effective mass is larger than the lightest active neutrino mass \cite{Fujii:2002jw}
\begin{align}\label{eq:lightest}
    \tilde{m}_1 > \text{Min}\left[m_\nu\right].
\end{align}
 We find that the $N_1$ decays after it has become non-relativistic ($K_1\ll1$) as long as 
\begin{align}\label{eq:param1}
    \tilde{m}_1 \ll \SI{2e-3}{\electronvolt}.
\end{align}
The energy density of the non-relativistic RHN redshifts slower than radiation, so it overtakes the radiation component and becomes the dominant contribution to the energy budget of the universe at  \cite{Giudice:1999fb}
\begin{align}
    T_\text{dom.} = \frac{7}{4} \frac{M_1}{g_*(T_\text{dom.})}\simeq 2\%\; M_1,
\end{align}
where we used that the number of relativistic degrees of freedom above the electroweak crossover is $g_*(T_\text{dom.})=\mathcal{O}(100)$. Once $\Gamma_1 = H(T_\text{dec.})$ the intermediate epoch of matter domination ends and the decays of $N_1$ to relativistic particles begin a new epoch of radiation domination with a starting temperature of 
\begin{equation}\label{eq:Tdec}
    T_\text{dec.} = \SI{3e8}{\giga\electronvolt}\sqrt{\frac{\tilde{m}_1}{10^{-6}\;\text{eV}}}\; \left(\frac{M_1}{10^{10}\;\text{GeV}}\right) \left(\frac{106.75}{g_*(T_\text{dec.})}\right)^\frac{1}{4}.
\end{equation}  
The decay takes place after the onset of early matter domination for \cite{Giudice:1999fb}
\begin{align}\label{eq:param2}
    \tilde{m}_1 < \SI{2.9e-7}{\electronvolt}.
\end{align}
If $\tilde{m_1}$ is larger than this number, there will be no era of intermediate RHN matter domination and consequently the decays of the $N_1$ will not produce enough entropy to lead to an appreciable dilution of the inflationary tensor mode background (see the following discussion in section \ref{eq:sec}). This bound implies together with \eqref{eq:lightest} that the lightest active neutrino mass has to be smaller than $\SI{2.9e-7}{\electronvolt}$ meaning that for normal-ordering (NO) we consider the following neutrino spectrum \cite{ParticleDataGroup:2022pth}
\begin{align}\label{eq:NO}
     m_1 \simeq 0,\quad 
     m_2 \simeq \sqrt{\Delta m_\text{sol.}^2}\simeq \SI{8.6e-3}{\electronvolt},\quad 
      m_3 \simeq \sqrt{\Delta m_\text{sol.}^2+\Delta m_\text{atm.}^2}\simeq \SI{0.05}{\electronvolt}.
\end{align}
For the inverted ordering (IO) we would instead have a quasi-degenerate spectrum \cite{ParticleDataGroup:2022pth}
\begin{align}\label{eq:IO}
      m_1 \simeq \sqrt{|\Delta m_\text{sol.}^2+\Delta m_\text{atm.}^2|}\simeq \SI{0.0492}{\electronvolt}, \quad 
     m_2 \simeq \sqrt{|\Delta m_\text{atm.}^2|}\simeq \SI{0.05}{\electronvolt},\quad 
     m_3 \simeq 0.
\end{align}
Above we used the results of the global fit to neutrino oscillation data \cite{Esteban:2020cvm} including the atmospheric data from Super-Kamiokande \cite{Super-Kamiokande:2005wtt,Super-Kamiokande:2004orf}:
\begin{align}
    \text{NO:}& \quad \Delta m_\text{sol.}^2 = 7.42^{+0.21}_{-0.20}\times 10^{-5}\;\text{eV},\quad \Delta m_\text{atm.}^2 = 2.517^{+0.026}_{-0.028}\times 10^{-3}\;\text{eV},\\
    \text{IO:}&  \quad \Delta m_\text{sol.}^2 = 7.42^{+0.21}_{-0.20}\times 10^{-5}\;\text{eV},\quad \Delta m_\text{atm.}^2 = -2.498^{+0.028}_{-0.028}\times 10^{-3}\;\text{eV}.
\end{align}
The duration of the intermediate matter dominated era can be expressed in terms of the number of $e$-foldings
\begin{align}
    N_e &= \text{log}\left(\frac{a(T_\text{dec.})}{a(T_\text{dom.})}\right)\simeq  \text{log}\left(\frac{25.4}{g_*(T_\text{dom.})} \left(\frac{v^2}{\tilde{m}_1 M_\text{Pl.}}\right)^\frac{2}{3}\right),\\
        &\simeq \begin{cases}
            & 0.3 \quad \text{for} \quad \tilde{m}_1 =\SI{2e-7}{\electronvolt},\\
            & 5\;\;\; \quad \text{for}  \quad \tilde{m}_1 =\SI{2e-10}{\electronvolt},
        \end{cases}
\end{align}
where we used that during matter domination $a\sim H^{-2/3}$ together with $H(T_\text{dec.})=\Gamma_1$ and $H(T_\text{dom.})\sim T_\text{dom.}^2/M_\text{Pl.}$ at the transition from radiation to matter domination.\\
Throughout this work we assume an initial equilibrium distribution for $N_1$. For small Yukawa couplings giving rise to  $\tilde{m_1}<10^{-3}\;\text{eV}$ \cite{Giudice:1999fb} the interactions in \eqref{eq:Seesaw} do not suffice to establish equilibrium in the radiation dominated plasma after inflationary reheating at $T_\text{RH}$. Hence our scenario precludes thermal leptogenesis and is sensitive to the initial conditions of the radiation bath. This is why we assume the initial population of RHN is produced by additional interactions such as couplings to the inflaton  $\varphi$   \cite{Hahn-Woernle:2008tsk} like \textit{e.g.}
\begin{align}
Y_{\varphi N} \;\varphi \overline{N^c} N,
\end{align}
for a production during reheating, or new gauge bosons from \textit{e.g.} GUTs \cite{Fritzsch:1974nn,Georgi:1974my} or gauged B-L \cite{Bezrukov:2009th}. Concentrating on the case of a $\text{U}(1)_\text{B-L}$ gauge boson with mass $m_{Z'}=g_\text{B-L} v_\text{B-L}>T_\text{RH}$ as an example, the scattering rate of $N_1$ with the SM quarks and leptons via off-shell $Z'$ would read approximately
\begin{align}
    \Gamma_\text{scat.}\simeq \frac{g_\text{B-L}^4 T^5}{m_{Z'}^4}=  \frac{T^5}{v_\text{B-L}^4}.
\end{align}
This interaction freezes-out while the $N_1$ are still relativistic ($T_\text{FO}>10 M_1$) as long as
\begin{align}\label{eq:relvev}
    v_\text{B-L}> \SI{7e11}{\giga\electronvolt}\cdot \left(\frac{M_1}{\SI{7.5e8}{\giga\electronvolt}}\right)^\frac{3}{4}\cdot \left(\frac{106.75}{g_{*\rho}(T_\text{FO})}\right)^\frac{1}{8}.
\end{align}
The impact of the underlying $\text{U}(1)_\text{B-L}$ breaking on stochastic GWs is briefly explained in section \ref{sec:other}.

\subsection{Non-thermal leptogenesis}\label{sec:lepto}
\noindent   We assume the inflationary reheating dynamics satisfy $M_2,M_3>T_\text{max}>M_1$ so that we can focus on the decays of the lightest RHN  $N_1$. In this context we defined $T_\text{max}>T_\text{RH}$ as the largest temperature during the epoch of inflationary reheating \cite{Garcia:2017tuj,Garcia:2020eof,Datta:2022jic}, which ends with a radiation bath of the temperature $T_\text{RH}$.  
Alternatively, if one assumes only $M_2,M_3\gtrsim (3-10) \times M_1$, the population of $N_{2,3}$ will have decayed away long before $N_1$ decays, as a consequence of their larger Yukawa couplings needed to explain the observed neutrino masses. Further we assume there is no primordial lepton asymmetry \textit{e.g.} from the decays of $N_{2,3}$.
Since the $N_1$ are too weakly coupled, they would not be able to erase this preexisting asymmetry \cite{Engelhard:2006yg}. However for realistic light neutrino masses the $N_{2,3}$ will be in the strong washout regime $\tilde{m}_{2,3}>10^{-3}\;\text{eV}$, so that  inverse decays  $L H \rightarrow N_{2,3}$ destroy a large portion of the asymmetry produced by the decays of $N_{2,3}$. 
The lepton asymmetry $n_\text{B-L} /s$, defined in terms of the number density of leptons minus anti-leptons normalized to the entropy density $s$, can be converted into a baryon asymmetry via the electroweak sphaleron process. For the RHN dominated scenario one finds a baryon asymmetry of  \cite{Giudice:1999fb}
\begin{equation}
   \frac{n_\text{B}}{s} = \frac{3}{4}\;c_\text{sph.}\cdot \varepsilon_1 \cdot  \frac{T_\text{dec.}}{M_1} \cdot \omega.
\end{equation}
The parameter $\varepsilon_1$   denotes the CP-violating decay parameter encoding the amount of leptonic asymmetry produced per decay of $N_1$. The sphaleron redistribution coefficient is found to be $c_\text{ph.}= 28/79$ \cite{PhysRevD.42.3344} and the term $\omega$, that will be determined later in this paragraph, parameterizes the washout of the lepton asymmetry.
Our analysis is different from the more commonly studied case of non-thermal leptogenesis immediately after inflationary reheating \cite{Lazarides:1990huy,Asaka:2002zu}, where $T_\text{dec.}/M_1$ would have to be replaced with $T_\text{RH}/m_\varphi$ with $m_\varphi$ being the inflaton mass, because here the RHN decay takes place much later, after it had time to dominate the energy budget  of the universe.
The factor of $T_\text{dec.}/M_1< 2\%$ comes from $n_{N}/s$, which can be  obtained from energy conservation  ($\rho_\text{tot.}= M_1 n_N$ before the decay) leading to 
\begin{equation}
    n_N = \frac{\pi^2}{30} g_{*\rho}(T_\text{dec.}) \frac{T_\text{dec.}^4}{M_1}
\end{equation}
and can be understood as the entropy dilution from the $N_1$ reheating:
The dimensionless dilution factor from the entropy produced by the instantaneous\footnote{Reference \cite{Ertas:2021xeh} goes beyond this approximation and also deals with  the case of a decaying particle whose temperature is different from the SM bath.} out-of-equilibrium decay of the dominating RHN $N_1$ \cite{PhysRevD.31.681,Kolb:1990vq,Bezrukov:2009th}   reads
 \begin{align}
    \Delta\equiv \frac{s(T_\text{dec.})a^3(T_\text{dec.})}{s(T_\text{RH})a^3(T_\text{RH})} &= \left(1+2.95 \left(\frac{2\pi^2 \braket{g_*(T)}}{45}\right)^\frac{1}{3} \frac{\left(\frac{n_N^i}{s} M_1\right)^\frac{4}{3}   }{\left(M_\text{Pl.}\Gamma_1\right)^\frac{2}{3}}  \right)^\frac{3}{4}\label{eq:Sanaly}\\
  \quad (\text{for}\quad \Delta\gg1) \quad   &\simeq 18.4\cdot \sqrt{\frac{10^{-10}\;\text{eV}}{\tilde{m_1}}} \left(\frac{106.75}{g_*(T_\text{dec.})}\right)^\frac{3}{4}\label{eq:delta} .
 \end{align}
In this context we denote the average of $g_*(T)$ over the decay period as $\braket{g_*(T)}$ and we assume that $\braket{g_*(T)}\simeq g_*(T_\text{dec.})$.  To obtain the second line we assumed for the initial abundance $n_N^i / s$ that $N_1$ decoupled from the plasma while relativistic to maximize the amount of entropy produced \cite{Bezrukov:2009th}, see also \eqref{eq:relvev}.
For hierarchical RHN spectrum ($M_3>M_2>M_1$) the decay parameter from the interference between tree-level and one-loop vertex- and self-energy-corrections is found to be \cite{Hambye:2003rt}
\begin{align}\label{eq:eps-gen}
    |\varepsilon_1|^\text{hier.} =  \sum_{i\neq 1} \frac{3}{16\pi} \frac{M_1}{M_i} \frac{\text{Im}\left(\left(\lambda\lambda^\dagger\right)^2_{1i}\right)}{\left|\lambda\lambda^\dagger\right|_{11}} =  \frac{3}{16\pi} \frac{M_1}{v^2}\frac{\sum_{i} m_i^2 \text{Im}\left(R_{1i}^2\right)}{\sum_{j} m_j |R_{1j}|^2 }<\varepsilon_\text{max},
\end{align}
where the upper limit (for normal ordered neutrino masses) reads  \cite{Davidson:2002qv}
\begin{align}\label{eq:max}
   \varepsilon_\text{max}= \frac{3}{16\pi }\frac{M_1}{v^2} (m_3-m_1).
\end{align}
It is worth mentioning that while the small required value of $\tilde{m}_1$ in \eqref{eq:Tdec} necessitates  small values of $|R_{1i}|^2$, this does not automatically force $  |\varepsilon_1|^\text{hier.}$ to be tiny, since this quantity depends only on a ratio of squared $R$-matrix elements. For completeness let us mention that for a  degenerate spectrum with $M_3 > M_2\simeq M_1$ the self-energy graph gets resonantly enhanced and the estimate gets modified as \cite{Hambye:2003rt}
\begin{align}
    |\varepsilon_1|^\text{degen.} = \varepsilon_\text{max} \cdot \frac{S_2 \cdot m_3-m_1}{m_3-m_1},\quad \text{where }\quad  
  S_2 \equiv \frac{M_2}{2 \Gamma_2}\quad \text{as long as} \quad M_2-M_1 = \frac{\Gamma_2}{2}.
\end{align}
We estimate the baryonic asymmetry for a general value of $\varepsilon_1$
\begin{align}
    \frac{n_\text{B}}{s} &\simeq 0.15  \cdot \frac{\sqrt{\tilde{m}_1 M_\text{pl.}}}{v}\cdot \varepsilon_1\cdot \omega, \\
    &\simeq 8.75\times 10^{-11} \cdot \sqrt{\frac{\tilde{m}_1}{\SI{2e-7}{\electronvolt}}}\cdot \left(\frac{\varepsilon_1 \cdot \omega}{2.4\times10^{-8}}\right),\label{eq:B-L}
\end{align}
where we chose $\tilde{m}_1$ for matter domination according to \eqref{eq:param2}.
One can compute the observed $n_B / s$ from the baryon-to-photon-ratio in \eqref{etaBobs} by making use of $s\simeq 7.04\; n_\gamma$. The required mass $M_1$ for the hierarchical spectrum can be obtained from \eqref{eq:max}
\begin{align}\label{eq:M1}
    M_1 \gtrsim \SI{2.44e8}{\giga\electronvolt}\cdot \left(\frac{n_\text{B-L} / s }{8.75\times10^{-11}} \right)\cdot \sqrt{\frac{\SI{2e-7}{\electronvolt}}{\tilde{m}_1}}  \cdot \left(\frac{\SI{0.05}{\electronvolt}}{m_3-m_1}\right)\cdot \left( \frac{1}{\omega}\right)
\end{align}
and depends intimately on the details of the active neutrino mass spectrum.
Note that unlike the usual Davidson-Ibarra bound $M_1\gtrsim 10^9\;\text{GeV}$ \cite{Davidson:2002qv} our estimate depends on the parameter $\tilde{m}_1$  due to the entropy produced in the RHN decay. It is not surprising that this bound can be slightly lower than the Davidson-Ibarra limit, as the out-of-equilibrium RHN abundance at $T_\text{dec.}$ can be larger than the typically assumed relativistic thermal yield. Fitting $M_1, \tilde{m}_1$ to the baryon asymmetry of the universe leads to $T_\text{dec.} \gtrsim \SI{3.3e6}{\giga\electronvolt}$ \cite{Hamaguchi:2001gw} and the condition $M_1>T_\text{dec.}$ is always satisfied for the range of $\tilde{m}_1$ we consider (see the discussion above \eqref{eq:param2}). It is important to point out that our present treatment ignores flavour effects \cite{Nardi:2005hs,Nardi:2006fx,Abada:2006ea,Abada:2006fw} such as the charged lepton Yukawa interactions being fast compared to the Hubble scale at different temperatures. These effects can change the asymmetry and consequently the Davidson-Ibarra bound by order one numbers \cite{Abada:2006fw} and are expected to be most relevant in the strong washout regime $\tilde{m}_1>10^{-3}\;\text{eV}$ \cite{Nardi:2006fx} not applicable here. Now let us take into account the washout of the asymmetry instantaneously produced at $T_\text{dec.}$. Because the universe transitions back to a second phase of radiation domination at $T_\text{dec.}$, we can reuse the standard estimates for washout. Since the inverse decay requires an on-shell $N_1$ it gets Boltzmann-suppressed and scales as \cite{Buchmuller:2004nz}
\begin{align}
    \Gamma_\text{ID} \sim \Gamma_1 e^{-\frac{M_1}{T}}.
\end{align}
Consequently   for  $T<T_\text{dec.}<M_1$ we can neglect the washout from inverse decays. That leaves the scattering processes $L L \leftrightarrow H^\dagger H^\dagger $ and $L H \leftrightarrow \overline{L} H^\dagger $ via intermediate RHNs $N_j\; (j=1,2,3)$.
Here one does not include the resonant contribution from on-shell $N_1$, as they are already included in the decay term of the Boltzmann equations \cite{Giudice:2003jh} and the masses of $N_{2,3}$ are not kinematically accessible. For  $T\ll M_1$  the scattering term can be expressed as \cite{Buchmuller:2004nz}
\begin{align}
    \Delta W \equiv \frac{2\times 10^{-6}}{z^2}\cdot \left(\frac{M_1}{\SI{2.5e8}{\giga\electronvolt}}\right)\cdot \left(\frac{\overline{m_\nu}}{\SI{0.05}{\electronvolt}}\right),
\end{align}
where  
\begin{align}
    z \equiv \frac{M_1}{T},\quad \text{and}\quad \overline{m_\nu} \equiv  \sqrt{3m_1^2 + 2 \Delta m_\text{sol.}^2 + \Delta m_\text{atm.}^2}
\end{align}
implying 
\begin{align}
    \omega &\simeq \exp\left(-\int_{z_\text{dec.}}^{\infty}\text{d}z\;\Delta W\right) \\
    &\simeq  \text{exp}\left(-2.7\times 10^{-9}\cdot \left(\frac{M_1}{\SI{2.5e8}{\giga\electronvolt}}\right)\cdot \left(\frac{\overline{m_\nu}}{\SI{0.05}{\electronvolt}}\right) \cdot \sqrt{\frac{\tilde{m}_1}{\SI{2e-7}{\electronvolt}}}\right).
\end{align}
In the above we  used equations \eqref{eq:NO} and \eqref{eq:IO} for the sum of neutrino masses $\overline{m_\nu}$. This process is negligible, if the absolute value of the exponent is  $\lesssim 0.1$  \cite{Hugle:2018qbw}, which corresponds to the bound 
\begin{align}
    M_1 < \SI{9e15}{\giga\electronvolt}\cdot \left(\frac{\SI{0.05}{\electronvolt}}{\overline{m_\nu}}\right)\cdot \sqrt{\frac{\SI{2e-7}{\electronvolt} }{\tilde{m}_1}},
\end{align}
compatible with the findings of \cite{Giudice:2003jh}, indicating that our parameter space (see \eqref{eq:M1}) will be save from any kind of washout: $\omega \simeq 1$.

\subsection{Dark Matter and Dark Radiation  Co-genesis}\label{sec:DM}

\noindent 
Dark Matter  could be included in Seesaw models via a lightest RHN with keV-scale masses \cite{Asaka:2005an,Asaka:2005pn} produced via either active-to-sterile oscillations \cite{Dodelson:1993je,Shi:1998km} or gauge interactions \cite{Bezrukov:2009th}.
The neutrino mass mediated by a keV-scale $N_1$ as DM is expected to be smaller than $\mathcal{O}\left(10^{-5}\;\text{eV}\right)$ \cite{Asaka:2005an}. 
Since then  $N_2$ would have  to play the role of the decaying particle for leptogenesis and we would have to require the associated effective neutrino mass to be below  $\mathcal{O}\left(10^{-7}\;\text{eV}\right)$ for matter domination (see \eqref{eq:param2}), we would not be able to explain both of  the observed neutrino mass splittings in \eqref{eq:NO} and \eqref{eq:IO}. Consequently we consider an additional particle as the DM.
The out-of-equilibrium decay  of a heavy $N_1$ to this particle might then  populate the dark matter abundance. A schematic model for this purpose consists of adding a gauge singlet Majorana fermion $\psi$ and a real singlet scalar $\sigma$, either of which (or both) could play the role of dark matter a priori.
This approach was first considered in reference \cite{Falkowski:2011xh} for the context of asymmetric dark matter and later in \cite{Falkowski:2017uya} for the case of CP-conserving decays to DM. The relevant couplings are
\begin{align}
    \mathcal{L}\supset   y\; N \sigma \psi + m_\psi\; \overline{\psi^c}\psi + V(H,\sigma).
\end{align}
For the sake of minimality we assumed that $\psi$ is a Majorana fermion. It might as well be a Dirac fermion, if we were to introduce a vector-like partner for it. We assume a general renormalizable scalar potential $ V(H,\sigma)$ for the real scalar $\sigma$ and that $M_1 \gg m_\psi + m_\sigma$. Additionally all portal couplings are presumed to be small enough to prevent thermal abundances of $\psi, \sigma$ in the early universe. The decay width of $N_1$ to $\psi\sigma$ reads
\begin{align}
    \Gamma_\psi \equiv    \Gamma(N_1\rightarrow \psi \sigma) = \frac{\left|y y^\dagger\right|_{11}}{16\pi} M_1,
\end{align}
where the factor of $1/2$ compared to \eqref{eq:dec} arises because this decay has singlets and not doublets in the final state. We define 
\begin{align}\label{eq:BR}
    \text{BR}_\psi = \frac{\Gamma_\psi}{\Gamma_1+\Gamma_\psi}\quad  \text{and} \quad \text{BR}_L = \frac{\Gamma_1}{\Gamma_1+\Gamma_\psi}.
\end{align}
The discussion in section \ref{sec:long} assumed that $\Gamma_1$ was the leading decay mode of $N_1$ determining the temperature $T_\text{dec.}$ at the end of the matter dominated phase  in \eqref{eq:Tdec}. Generally speaking this temperature should be calculated from $\text{Max}\left[\Gamma_1,\Gamma_\psi\right]$ instead. In order to use the parameter region from section \ref{sec:long} we will set $\text{BR}_L\geq\text{BR}_\psi$. In the following we will assume that $\psi$ is the DM, because  as long as $\sigma$ does not receive a vev \cite{Falkowski:2011xh} it has only a suppressed decay mode to $\nu_L \sigma$  for $m_\psi>m_\sigma$ via $\nu_L - N$ mixing, that will be discussed in a moment. Its yield is different from the typical Freeze-in approach \cite{Hall:2009bx,Liu:2020mxj} since the decaying RHN is not in thermal equilibrium with the rest of the bath anymore. It also differs from the  super-WIMP \cite{Feng:2003xh}, because the RHN is relativistic at decoupling unlike the non-relativistic WIMP that decays to DM. For our case one finds \cite{Kawasaki:1995cy,Gelmini:2006pw}
\begin{align}
    \frac{n_\psi}{s}= \text{BR}_\psi\; \frac{n_N}{s} = \frac{3}{4}\; \text{BR}_\psi\; \frac{T_\text{dec.}}{M_1},
\end{align}
from which we deduce that
\begin{align}
   \Omega_\psi h^2 \simeq 0.12 \;\cdot\left(\frac{m_\psi}{\SI{170}{\kilo\electronvolt}}\right)\cdot \left(\frac{\text{BR}_\psi}{5\times 10^{-4}}\right)\cdot\sqrt{\frac{\tilde{m}_1}{\SI{2e-7}{\electronvolt}}}\label{eq:DM}.
\end{align}
One can see that the DM abundance only constraints the product $m_\psi \text{BR}_\psi$
and we use it as a free parameter in the upcoming sections about gravitational waves instead of just $m_\psi$. For small branching fractions our scenario leads to light dark matter. Fermionic DM is only gravitationally bound to the DM halo of our galaxy if $m_\psi \gtrsim \mathcal{O}(\SI{100}{\electronvolt})$ \cite{PhysRevLett.42.407}. In order to comply with bounds from structure formation, that constrain the free-streaming scale of dark matter, we have to demand that \cite{Decant:2021mhj}
\begin{align}\label{eq:FS}
    m_\psi \gtrsim \mathcal{O}(\SI{10}{\kilo\electronvolt}).
\end{align}

\begin{figure}[t]
    \centering   
    \includegraphics[width=0.6\textwidth]{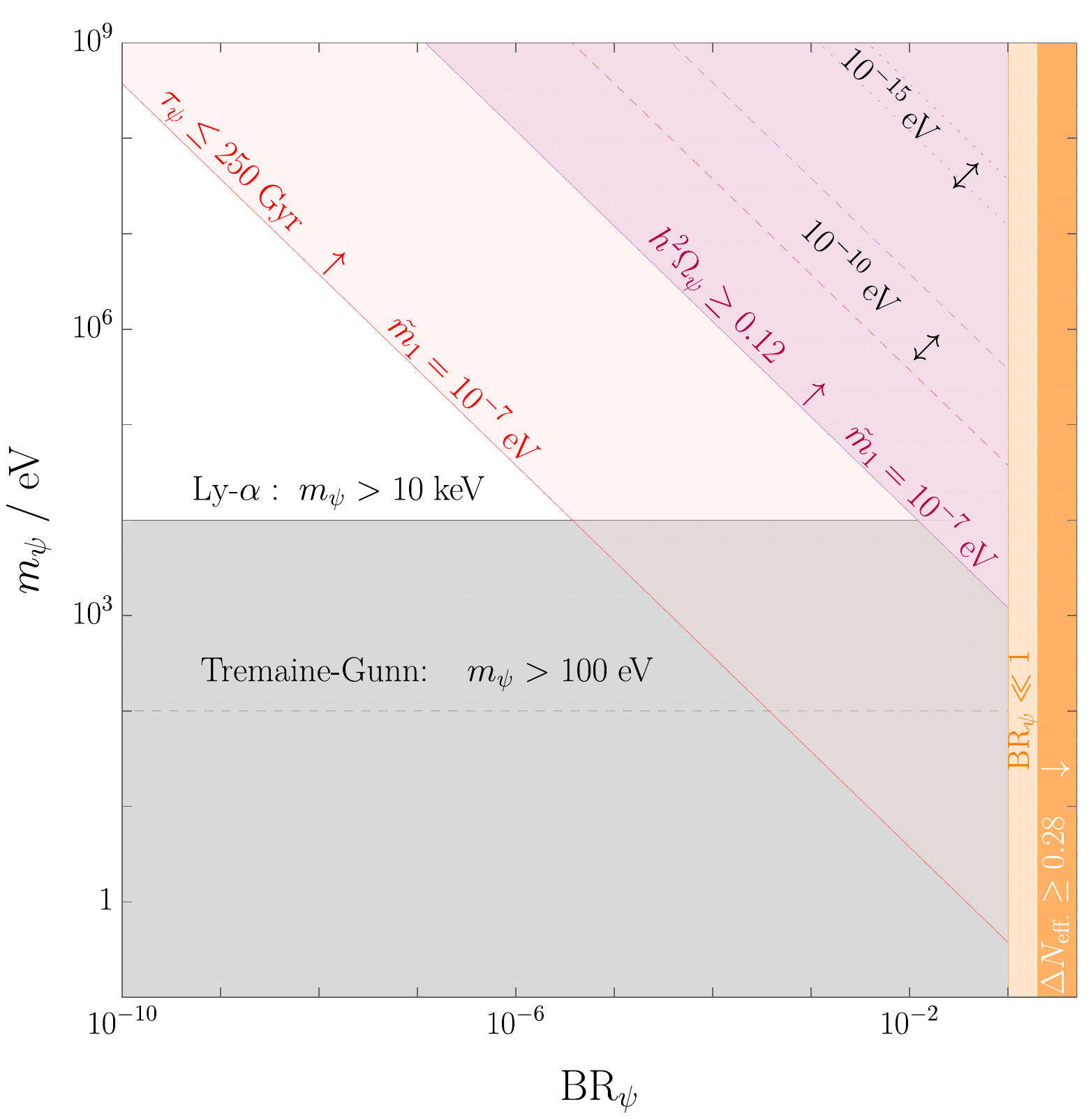}
    \caption{Parameter space for the dark matter mass $m_\psi$ versus the branching ratio $\text{BR}_\psi$ of the RHN decay to dark matter. Contours with (straight, dashed, dotted) lines correspond to $\tilde{m}_1 = \left(10^{-7},10^{-10},10^{-15}\right)\;\text{eV}$ . The purple contours reproduce the observed dark matter relic abundance and above the contour the abundance would be too large (for fixed $\tilde{m}_1$). The gray regions are excluded because of unsuccessful structure formation (Lyman-$\alpha$) and dark matter not being gravitationally bound (Tremaine-Gunn). On the red contours for    the DM lifetime from $\psi\rightarrow \nu_L \sigma$  is  equal to the   observational limit and in the colored region above  (for fixed $\tilde{m}_1$) the lifetime would be too small.
    This  excludes the lines with $\tilde{m}_1 = \left(10^{-7},10^{-10}\right)\;\text{eV}$, meaning that here only $\tilde{m}_1 =10^{-15}\;\text{eV}$ is viable for DM. Note that lifetime bound   disappears for  $m_\psi <m_\sigma$, in which case the entire purple region is allowed.  The  area in light orange is excluded by our assumption $\text{BR}_\psi \ll \text{BR}_L \simeq 1$ and the orange region   would be  
    excluded, if the real scalar also produced in the RHN decay was stable and light enough to be dark radiation (see the discussion below \eqref{eq:neff2}).}
    \label{fig:DM}
\end{figure}

\noindent Both of these constraints illustrate why we need $ \text{BR}_\psi \ll  \text{BR}_L$, which translates to $y_1\ll \lambda_{1i}$. In the regime $m_\sigma < m_\psi$ the following decay from $\nu_L - N_{1,2,3}$ mixing after electroweak symmetry breaking  becomes kinematically allowed \cite{Coy:2021sse} and we assume that  $m_\sigma \ll m_\psi$:
\begin{align}
    \Gamma\left(\psi \rightarrow \nu_L \sigma \right) = \frac{\left|y y^\dagger\right|_{11}}{16\pi} \sum_{i,j} \frac{\lambda_{ji} \lambda^\dagger_{ij}\; v^2}{M_j^2} m_\psi
   \simeq \frac{m_\psi\; \text{BR}_\psi}{8\pi} \frac{\tilde{m}_1 \sum_i m_i }{v^2} \frac{M_1}{M_{2,3}}
\end{align}
Here we summed over the final state lepton flavors, which together with the sum over all three  RHNs and making the approximation of factoring out one power of $M_{2,3}$, allows us to trade the $\lambda$-couplings of the active neutrino masses via the Seesaw-relation \eqref{eq:seesaw}. Equation \eqref{eq:BR} lets us trade the $y$-couplings for $\text{BR}_\psi$ and $\tilde{m}_1$ in the limit $ \text{BR}_\psi \ll  \text{BR}_L$. Data on baryon acoustic oscillations
and structure formation requires a lifetime  $\tau_\psi = 1/\Gamma\left(\psi \rightarrow \nu_L \sigma\right)$ for DM decaying to dark radiation of $\tau_\psi>(249.6-268.8)\times 10^9\;\text{yr}$ \cite{Simon:2022ftd} depending on the exact dataset used. The resulting bound for $\tau_\psi>250\times 10^9\;\text{yr}$ reads
\begin{align}
    m_\psi \text{BR}_\psi < \SI{1.8e-2}{\electronvolt} \cdot \left(\frac{\SI{2e-7}{\electronvolt}}{\tilde{m}_1  }\right)\cdot \left(\frac{\SI{0.05}{\electronvolt}}{\sum_i m_i}\right)\cdot \left(\frac{M_{2,3}/M_1}{3}\right)
\end{align}
and is compatible with the relic density \eqref{eq:DM} for
\begin{align}\label{eq:bound}
    \tilde{m}_1 < \SI{9.7e-15}{\electronvolt}\cdot \left(\frac{0.12}{ \Omega_\psi h^2}\right)^2.
\end{align}
We depict the allowed parameter space in figure \ref{fig:DM}.
Once can see that the parameters $\tilde{m}_1 = \left(10^{-7},10^{-10}\right)\;\text{eV}$ violate the lifetime constraint, because for each constant $\tilde{m}_1$ the purple relic abundance iso-contour line  is above the red line  for $\tau_\psi= 250\times 10^9\;\text{yr}$. The only viable parameter point in this plot has $\tilde{m}_1 = 10^{-15}\;\text{eV}$ in agreement with \eqref{eq:bound}, because here the lifetime iso-contour is above the line for the relic density and we find dark matter close to the GeV-scale. The previous limits only apply for $m_\psi >m_\sigma$. In general the scalar $\sigma$ couples  to the SM Higgs through the following terms 
\begin{align}\label{eq:portal}
     V(H,\sigma)\supset \lambda_{\sigma H} \sigma^2 \left|H\right|^2 + \left(\kappa\; \sigma +\text{h.c.}\right)  \left|H\right|^2.
\end{align}
For $m_\sigma >m_h$ it could decay to the SM Higgs.  If this is kinematically forbidden,  there  could be decay modes lighter SM fermions such as \textit{e.g.} the electron $\sigma \rightarrow e^+e^-e^+e^-$ via off-shell SM Higgs bosons. In case  $\sigma$ has no vev these decays require the $\kappa$ coupling. If $\sigma$ is too light to decay to SM states or the couplings  $\lambda_{H\sigma}$ and $\kappa$ are very small, then the relic abundance of $\sigma$ survives until today.
In this case and assuming that $\lambda_{H\sigma}$ and $\kappa$ are small enough to avoid thermalization with the SM plasma, the non-thermal $\sigma$ could still exist in the form of dark radiation. Its energy density is found from $n_\sigma = n_\psi = \text{BR}_\psi n_N$ to be \cite{Mazumdar:2016nzr}
\begin{align}
    \rho_\sigma(T_\text{dec.}) = \frac{ \pi^2}{30}g_{*}(T_\text{dec.})\; \text{BR}_\psi\; \frac{\sqrt{m_\sigma^2 + \left(\frac{M_1}{2}\right)^2}}{M_1}\;T_\text{dec.}^4
\end{align}
and we compute the abundance of dark radiation, conventionally parameterized as the number of additional neutrinos as \cite{Luo:2020fdt} assuming again that $M_1\gg m_\sigma$:
\begin{align}
    \Delta N_\text{eff.} &= \frac{4}{7}\cdot g_{*\rho}(T_\text{dec.})\cdot  \left(\frac{10.75}{g_{*S}(T_\text{dec.})}\right)^\frac{4}{3}\cdot \frac{\rho_\sigma(T_\text{dec.})}{\rho_\text{SM}(T_\text{dec.})}\\
    &\simeq 0.06 \cdot \left(\frac{\text{BR}_\psi}{4\%}\right) \cdot \left(\frac{106.75}{g_{*S}(T_\text{dec.})}\right)^\frac{4}{3}\cdot \left(\frac{g_{*}(T_\text{dec.})}{106.75}\right) \label{eq:neff2}
\end{align}
We see that $\sigma$ would lead to too much dark radiation compared with the current Planck bound $ \Delta N_\text{eff.}^\text{Planck+BAO} \simeq 0.28$ \cite{Planck:2018vyg} unless we make the branching ratio $\text{BR}_\psi$, which also controls the DM production,  smaller than about 20\% (see figure \ref{fig:DM}). However we saw previously that $\text{BR}_\psi$ can be far below a percent for heavy enough DM, which is why we do not necessarily expect observable dark radiation. BBN sets a bound of $\Delta N_\text{eff.}^\text{BBN}\simeq 0.4$ \cite{Cyburt:2015mya}.
The  projected sensitivities of upcoming experiments read
$\Delta N_\text{eff}^\text{proj.}=0.014$ for   CMB-HD  \cite{CMB-HD:2022bsz}, $\Delta N_\text{eff}^\text{proj.}=0.05$ for   CMB-Bharat  \cite{CMB-Bharat}, 
$\Delta N_\text{eff}^\text{proj.}=0.06$ for    CMB Stage IV \cite{Abazajian:2019eic,annurev-nucl-102014-021908} and NASA's PICO mission  \cite{NASAPICO:2019thw} or $\Delta N_\text{eff.} \lesssim 0.12$ for  CORE  \cite{CORE:2017oje},   the South Pole Telescope  \cite{SPT-3G:2014dbx} as well as the Simons observatory \cite{SimonsObservatory:2018koc}. Before closing let us emphasize again that $\sigma$ only counts as dark radiation when it is very light and stable or long-lived.

\section{Gravitational Waves}\label{sec:3}
\subsection{Distortion of the  inflationary tensor mode spectrum}\label{eq:sec}

\noindent
We assume primordial inflation ended in an epoch of reheating, creating a Standard Model plasma of radiation with an initial temperature $T_\text{RH}$ set by the reheating dynamics.
Gravitational waves produced during inflation first leave the horizon and have constant amplitudes while outside the horizon. After they re-enter the horizon the amplitude becomes damped.
The power spectrum of gravitational waves (GWs) today can be written as a function of the wave-number $k= 2\pi f$ with $f$ being the frequency
\begin{align}
    \Omega_\text{GW}(k) = \frac{1}{12}\left(\frac{k}{a_0 H_0}\right)^2 P_T(k),
\end{align}
where $a_0=1$ and $H_0\simeq \SI{2.2e-4}{\per\mega\parsec}$ \cite{Datta:2022tab} are the scale factor and expansion rate today and $P_T$ denotes the   spectrum of tensor modes. It is   parameterized in terms of the primordial power spectrum from inflation $P_T^\text{prim.}$
\begin{align}
    P_T(k) = T_T^2(k) \;P_T^\text{prim.}(k)
\end{align}
as well as a transfer function $T_T^2(k)$. This transfer function describes the propagation of GWs $h_{ij}$ in the Friedmann-Lemaitre-Robertson-Walker background 
\begin{align}
    h_{ij}''+2 a H h_{ij}' - \Delta h_{ij}=0,
\end{align}
where primes denote derivatives with respect to conformal time, after the horizon re-entry at a temperature of $T_\text{in}$  that  depends on the wave-number via  \cite{Nakayama:2008wy}
\begin{equation}
    T_\text{in} = \SI{5.8e6}{\giga\electronvolt}\cdot \left(\frac{106.75}{g_*(T_\text{in})}\right)^{\frac{1}{6}} \left(\frac{k}{\SI{e4}{\per\mega\parsec}}\right).
\end{equation}
The inflationary tensor power spectrum is conventionally parameterized in terms of its amplitude $A_T$ and its spectral index $n_T$ at the pivot scale $k_* =\SI{0.05}{\per\mega\parsec}$ \cite{Planck:2018jri}
\begin{align}
    P_T^\text{prim.}(k) = A_T(k_*) \left(\frac{k}{k_*}\right)^{n_T}.
\end{align}
This amplitude is related to the scalar power spectrum $P_\xi(k_*)= 2.0989\times 10^{-9}$ \cite{Planck:2018jri} via the tensor-to-scalar-ratio $r<0.035$ \cite{BICEP:2021xfz}
\begin{align}
    A_T(k_*) = r \; P_\xi(k_*).
\end{align} 
Observations of the cosmic microwave background only constrain the scalar spectral index to be $n_S=0.9649\pm0.0042$ \cite{Planck:2018jri}, which is why we take $n_T$ as a constant free parameter. The case of $n_T>0\; (<0)$ is known as a blue-tilted (red-tilted) spectrum. Standard single field slow-roll inflation predicts a red-tilted spectrum, as the tensor spectral index $n_T$
satisifes the so-called consistency relation  $n_T = - r /8$  \cite{Liddle:1993fq}, however this does not rule out the possibilities of a blue-tilted spectrum, which is well motivated in various scenarios including \textit{e.g.} string gas cosmology \cite{Brandenberger:2006xi}, super-inflation models \cite{Baldi:2005gk}, G-inflation \cite{Kobayashi:2010cm},
non-commutative inflation \cite{Calcagni:2004as,Calcagni:2013lya}, particle production during inflation \cite{Cook:2011hg,Mukohyama:2014gba}, and several others \cite{Kuroyanagi:2020sfw}. Here we will also seek to
investigate such scenarios from the perspective of models of the early universe and leptogenesis. An epoch of early or intermediate matter domination would change the transfer function compared to the standard case of radiation domination, and hence the expansion  of the background is imprinted in the damping of the gravitational wave amplitude.  References  \cite{Turner:1993vb,Chongchitnan:2006pe,Nakayama:2008wy,Nakayama:2009ce,Kuroyanagi:2011fy,Kuroyanagi:2014nba} computed this transfer function numerically and found a compact analytical expression   with  a fitting function $F(k)$
\begin{align}
    T_T^2(k) &= \Omega_m^2 \left(\frac{g_*(T_\text{in})}{g_*^0}\right)\left(\frac{g_{*S}^0}{g_{*S}(T_\text{in})}\right)^\frac{4}{3} \left(\frac{3j_1(z_k)}{z_k}\right)^2 F(k)
\end{align}
in terms of the total matter density $\Omega_m =0.31$, the first spherical Bessel function $j_1(z_k)$ and $z_k\equiv k\; \tau_0$ with $\tau_0=2 / H_0$ \cite{Datta:2022tab} being the conformal time today. The factors of the relativistic degrees of freedom encode the expansion of the universe and we use the  fitting functions of reference \cite{Kuroyanagi:2014nba} for $g_*(T_\text{in})$ and $g_{*S}(T_\text{in})$ with the present day values $g_*^0=3.36$ and $g_{*S}^0=3.91$, whereas the Bessel function describes the damping of the gravitational wave amplitude after horizon re-entry. In the limit $z_k\gg1$, which always holds for the frequencies we are interested in,  
\begin{align}
    k\; \tau_0 \simeq 6\times10^{15} \left(\frac{f}{10^{-3}\;\text{Hz}}\right),
\end{align}
we can trade the oscillatory $j_1(z_k)$ for $1/(\sqrt{2}z_k)$. Note that in  references \cite{Kuroyanagi:2011fy,Kuroyanagi:2014nba} the correct limiting behavior was mentioned for the wrong limit $z_k\ll 1$ (for which one would obtain $j_1(z_k)\sim z_k$ instead) .  We employ the most recent results of \cite{Kuroyanagi:2014nba} for the fitting function $F(k)$.
Without intermediate matter domination it reads
\begin{align}\label{eq:stand}
   F(k)_\text{standard}  =   T_1^2\left(\frac{k}{k_\text{eq.}}\right)T_2^2\left(\frac{k}{k_\text{RH}}\right),
\end{align}
whereas including an epoch of RHN domination leads to 
\begin{align}\label{eq:IMD}
    F(k)_\text{IMD} &=  
      T_1^2\left(\frac{k}{k_\text{eq.}}\right)T_2^2\left(\frac{k}{k_\text{dec.}}\right)T_3^2\left(\frac{k}{k_\text{dec. S}}\right)T_2^2\left(\frac{k}{k_\text{RH S}}\right).
\end{align}
Here we introduce
 \begin{align}
     k_\text{eq.} &= \SI{7.1e-2}{\per\mega\parsec}\cdot \Omega_m h^2, \\
     k_\text{dec.} &= \SI{1.7e14}{\per\mega\parsec} \left(\frac{g_{*S}(T_\text{dec.})}{g_{*S}^0}\right)^\frac{1}{6} \left(\frac{T_\text{dec.}}{10^7\;\text{GeV}}\right),\label{eq:kdec}\\
     k_\text{RH} &= \SI{1.7e14}{\per\mega\parsec} \left(\frac{g_{*S}(T_\text{RH})}{g_{*S}^0}\right)^\frac{1}{6} \left(\frac{T_\text{RH}}{10^7\;\text{GeV}}\right),\\
     k_\text{dec. S}&= k_\text{dec.} \Delta^\frac{2}{3},\\
     k_\text{RH S} &= k_\text{RH} \Delta^{-\frac{1}{3}},
 \end{align}
where all quantities with a subscript (superscript) \enquote{0} are evaluated today and we set $h=0.7$. The entropy dilution factor $\Delta$ was defined in \eqref{eq:Sanaly} and the fit functions read
 \begin{align}
     T_1^2(x) &= 1+1.57 x +3.42 x^2,\\
     T_2^2(x) &= (1-0.22x^\frac{3}{2}+0.65x^2)^{-1},\\
     T_3^2(x) &= 1+0.59 x +0.65 x^2.
 \end{align}
 Physically $T_1$ describes the transition from a radiation dominated phase to a matter dominated epoch and $T_2$ the case of going from matter domination to radiation domination. $T_3$ has the same physical interpretation as $T_1$ but allows for  a better numerical fit \cite{Kuroyanagi:2014nba}. 
One deduces from the wave-number $k_\text{dec.}= 2\pi  f_\text{sup.} $ at the time of RHN decay in \eqref{eq:kdec} that the gravitational wave spectrum gets suppressed by the entropy dilution for frequencies above  
\begin{align}
    f_\text{sup.}
    &\simeq 2.7\times10^{-10}\;\text{Hz}\; \left(\frac{T_\text{dec.}}{\SI{10}{\mega\electronvolt}}\right),\\
    &\simeq 9\times10^{-2}\;\text{Hz}\cdot \left(\frac{n_\text{B}/s}{\SI{8.75e-11}{}}\right) \cdot \left(\frac{\SI{0.05}{\electronvolt}}{m_3-m_1}\right)\cdot \left(\frac{106.75}{g_*(T_\text{dec.})}\right)^\frac{1}{4}\label{eq:freq},
\end{align}
where in the last line we fixed $M_1$  via equation \eqref{eq:M1} to reproduce the observed baryon asymmetry, which means that all the RHN decay at $T_\text{dec.} = \SI{3.3e6}{\giga\electronvolt}$ hence the constant  $ f_\text{sup.}$. The suppression factor of the power spectrum is  \cite{Seto:2003kc}
\begin{align}
    R_\text{sup.} = \frac{\Omega_\text{GW}^\text{IMD}}{\Omega_\text{GW}^\text{standard}}\simeq \frac{1}{ \Delta^{\frac{4}{3}}} \label{eq:dil},
\end{align}
which  depends only on $\tilde{m_1}$ via $\Delta$ in \eqref{eq:delta}.
Here   $\Omega_\text{GW}^\text{IMD}$ was computed from \eqref{eq:IMD} and takes the intermediate matter domination (IMD) from the RHN   into account, whereas $\Omega_\text{GW}^\text{standard}$ from \eqref{eq:stand} appears in the absence of RHN domination.

\subsection{Other GW sources }\label{sec:other}
So far, when it comes to gravitational waves, most studies involving the Seesaw mechanism  have focused on the dynamics of \textit{e.g.} the $\text{U}(1)_\text{B-L}$ breaking, which underlies the RHN Majorana masses in unified gauge theories  \cite{Fritzsch:1974nn,Georgi:1974my}. The dynamics of the scalar responsible for breaking this gauge symmetry can source a separate stochastic gravitational wave background by means of a first order \cite{Chao:2017ilw,Okada:2018xdh,Hasegawa:2019amx,Haba:2019qol} or second order \cite{Buchmuller:2013lra,Buchmuller:2019gfy,Dror:2019syi}   phase transition as well as via the formation of a network of cosmic strings \cite{PhysRevD.31.3052,Dror:2019syi,Blasi:2020wpy,Dunsky:2021tih} via the Kibble mechanism \cite{Kibble:1976sj}.  If the phase transition or the formation of topological defects happens before inflation - and the symmetry is never (non-)thermally restored -  any trace of the B-L transition will be diluted away due to the exponential expansion of space-time. The symmetry is broken throughout inflation and reheating if \cite{Redi:2022llj}
\begin{align}
    v_\text{B-L} > \text{Max}\left[\frac{H_I}{2\pi}, T_\text{max.}\right],
\end{align}
where the first term is the Gibbons-Hawkings temperature  \cite{PhysRevD152738}  in terms of the Hubble rate during inflation $H_I$ and the second term the maximum temperature  during reheating \cite{Garcia:2017tuj,Garcia:2020eof,Datta:2022jic}, which can be drastically larger than the temperature of the radiation bath at the end of reheating $T_\text{RH}$. Since $ T_\text{max.}$ depends on the reheating scenario, the best we can do to get an estimate on $v_\text{B-L}$ is to assume that  $H_I /(2\pi) >  T_\text{max.}$ and saturate the current CMB-limit on $H_I\lesssim \SI{2.5e14}{\giga\electronvolt}$ \cite{Akrami:2018odb} leading to
\begin{align}
    v_\text{B-L} \gtrsim \SI{4e13}{\giga\electronvolt}.
\end{align}
This further motivates why we consider high scale leptogenesis. Moreover this bound is compatible with the condition \eqref{eq:relvev} for a thermalized population of $N_1$ from B-L gauge scatterings.
Also note that one could even consider a case, where no additional degrees of freedom except the RHN are added to the SM below the Planck scale, so that there would be no source for the stochastic GW background (in this case the initial thermal RHN abundance would have to come from inflaton decays). Consequently our high scale scenario without a stochastic GW background, being essentially independent of the dynamics of the  $\text{U}(1)_\text{B-L}$ transition and the associated scalar, can be viewed as complementary to the existing analyses.

\subsection{Detectors and signal-to-noise ratio}
\noindent We  display the (expected) sensitivity curves for a variety of exisiting and proposed experiments that can be grouped in terms of 
\begin{itemize}
    \item \textbf{ground based  interferometers:} \textsc{LIGO}/\textsc{VIRGO}             \cite{LIGOScientific:2016aoc,LIGOScientific:2016sjg,LIGOScientific:2017bnn,LIGOScientific:2017vox,LIGOScientific:2017ycc,LIGOScientific:2017vwq}, a\textsc{LIGO}/a\textsc{VIRGO}  \cite{Harry_2010,LIGOScientific:2014pky,VIRGO:2014yos,LIGOScientific:2019lzm}, \textsc{AION} \cite{Badurina:2021rgt,Graham:2016plp,Graham:2017pmn,Badurina:2019hst}, \textsc{Einstein Telescope (ET)} \cite{Punturo:2010zz,Hild:2010id}, \textsc{Cosmic Explorer (CE)}  \cite{LIGOScientific:2016wof,Reitze:2019iox},
    \item   \textbf{space based interferometers:}  \textsc{LISA} \cite{2017arXiv170200786A,Baker:2019nia}, \textsc{BBO} \cite{Crowder:2005nr,Corbin:2005ny,Harry_2006}, 
    \textsc{DECIGO}, \textsc{U-DECIGO}\cite{Seto:2001qf,Kudoh:2005as,Kawamura_2006,Nakayama:2009ce,Yagi:2011wg,Kawamura:2020pcg}, \textsc{AEDGE} \cite{AEDGE:2019nxb,Badurina:2021rgt}, \textsc{$\mu$-ARES} \cite{Sesana:2019vho}
    \item \textbf{recasts of star surveys:} \textsc{GAIA}/\textsc{THEIA} \cite{Garcia-Bellido:2021zgu}, 
    \item \textbf{pulsar timing arrays (PTA):} \textsc{SKA} \cite{Carilli:2004nx,Janssen:2014dka,Weltman:2018zrl}, \textsc{EPTA} \cite{Kramer_2013,Lentati:2015qwp,Babak:2015lua}, \textsc{NANOGRAV}~\cite{McLaughlin:2013ira,NANOGRAV:2018hou,Aggarwal:2018mgp,Brazier:2019mmu,NANOGrav:2020bcs}
    \item \textbf{CMB polarization:} Planck 2018 \cite{Akrami:2018odb} and BICEP 2/ Keck \cite{BICEP2:2018kqh} computed by  \cite{Clarke:2020bil}, \textsc{LiteBIRD} \cite{Hazumi:2019lys}, 
    \item \textbf{CMB spectral distortions:} \textsc{PIXIE}, \textsc{Super-PIXIE}  \cite{Kogut_2011,Kogut:2019vqh}, \textsc{VOYAGER2050}~\cite{Chluba:2019kpb}
\end{itemize}

\noindent Interferometers measure displacements in terms of a so called dimensionless strain-noise $h_\text{GW}(f)$ that is related to the GW amplitude and can be converted into the corresponding  energy density \cite{Garcia-Bellido:2021zgu}
\begin{align}
    \Omega_\text{exp}(f) h^2 = \frac{2\pi^2 f^2}{3 H_0^2} h_\text{GW}(f)^2 h^2,
\end{align}
with $H_0 = h\times 100 \;\text{(km/s)}/\text{Mpc}$ being the Hubble rate today. 
We compute the signal-to-noise ratio (SNR) for a given or projected experimental sensitivity $\Omega_\text{exp}(f)h^2$ in order to assess the detection probability of the primordial GW background via the following prescription~\cite{Thrane:2013oya,Caprini:2015zlo}
\begin{align}
     \text{SNR}\equiv \sqrt{\tau \int_{f_\text{min}}^{f_\text{max}} \text{d}f \left(\frac{ \Omega_\text{GW}(f) h^2}{\Omega_\text{exp}(f) h^2}\right)^2 } \label{eq:SNR},
\end{align}
where $h=0.7$ and  $\tau = 4\; \text{years}$ is the observation time. For this analysis we consider $\text{SNR}\geq 10$ as the  detection threshold.

\subsection{Dark radiation bounds from BBN and CMB decoupling}
The energy density in gravitational waves  should be smaller than the limit on dark radiation encoded in $\Delta N_\text{eff.}$ from Big Bang Nucleosynthesis and CMB observations (see the discussion below \eqref{eq:neff2} for bounds and projections on  $\Delta N_\text{eff.}$) \cite{Maggiore:1999vm}
\begin{align}
    \int_{f_\text{min}}^{f=\infty} \frac{\text{d}f}{f}   \Omega_\text{GW}(f) h^2 \leq 5.6\times10^{-6}\;\Delta N_\text{eff.}.\label{eq:darkrad}
\end{align}
The lower limit of the integration is $f_\text{min}\simeq 10^{-10}\text{Hz}$ for BBN and  $f_\text{min}\simeq 10^{-18}\text{Hz}$ for the CMB. In practice, when \textit{e.g.} plotting many GW spectra simultaneously, and as a first estimate we  neglect the frequency dependence to constrain the energy density of the peak for a given spectrum 
\begin{align}
    \Omega_\text{GW}^\text{Peak} h^2 \leq   5.6\times10^{-6}\;\Delta N_\text{eff.}.\label{eq:darkrad2}
\end{align}

\begin{figure}[t]
    \centering
    \includegraphics[width=0.45\textwidth]{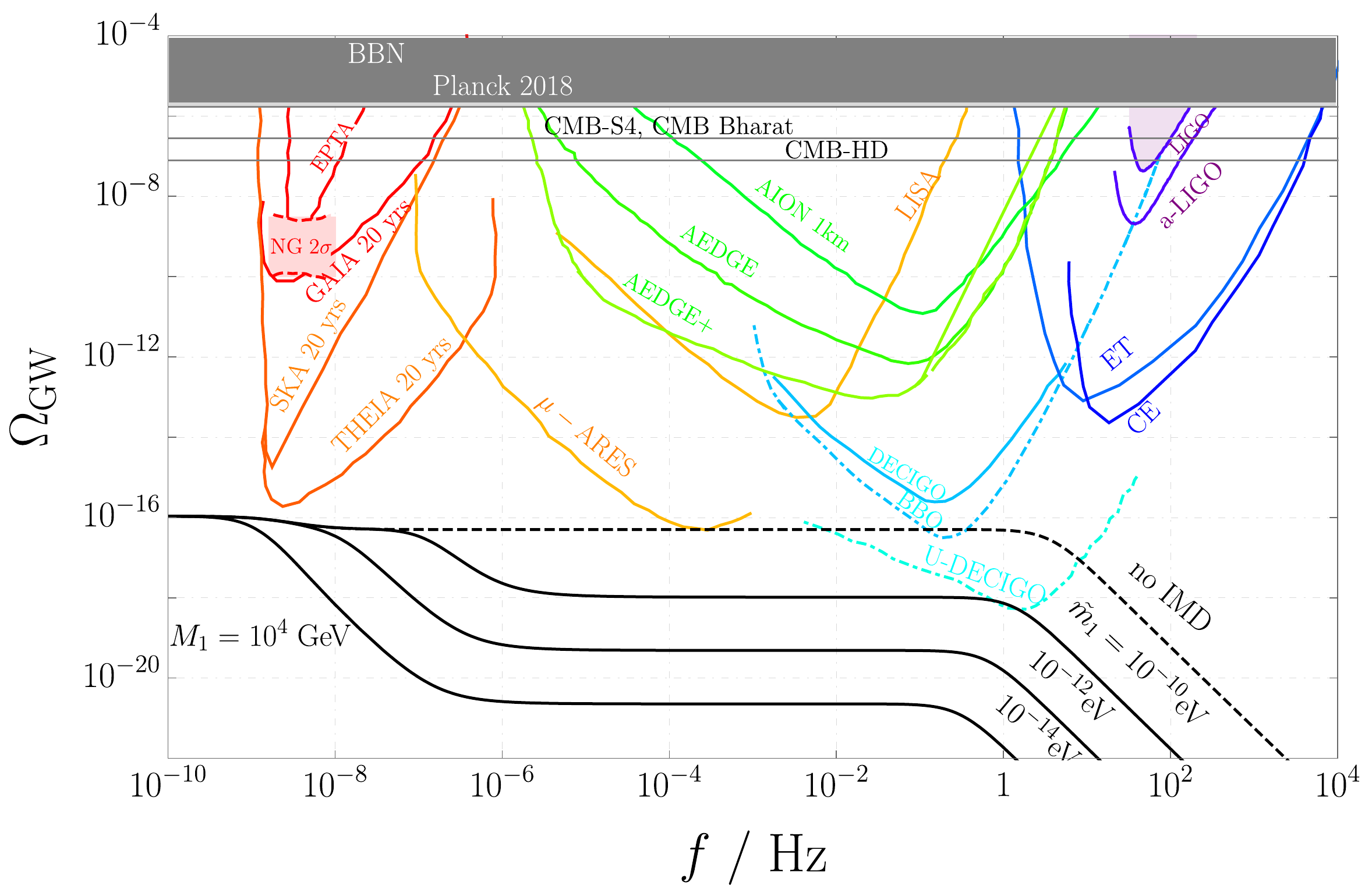}
    \includegraphics[width=0.45\textwidth]{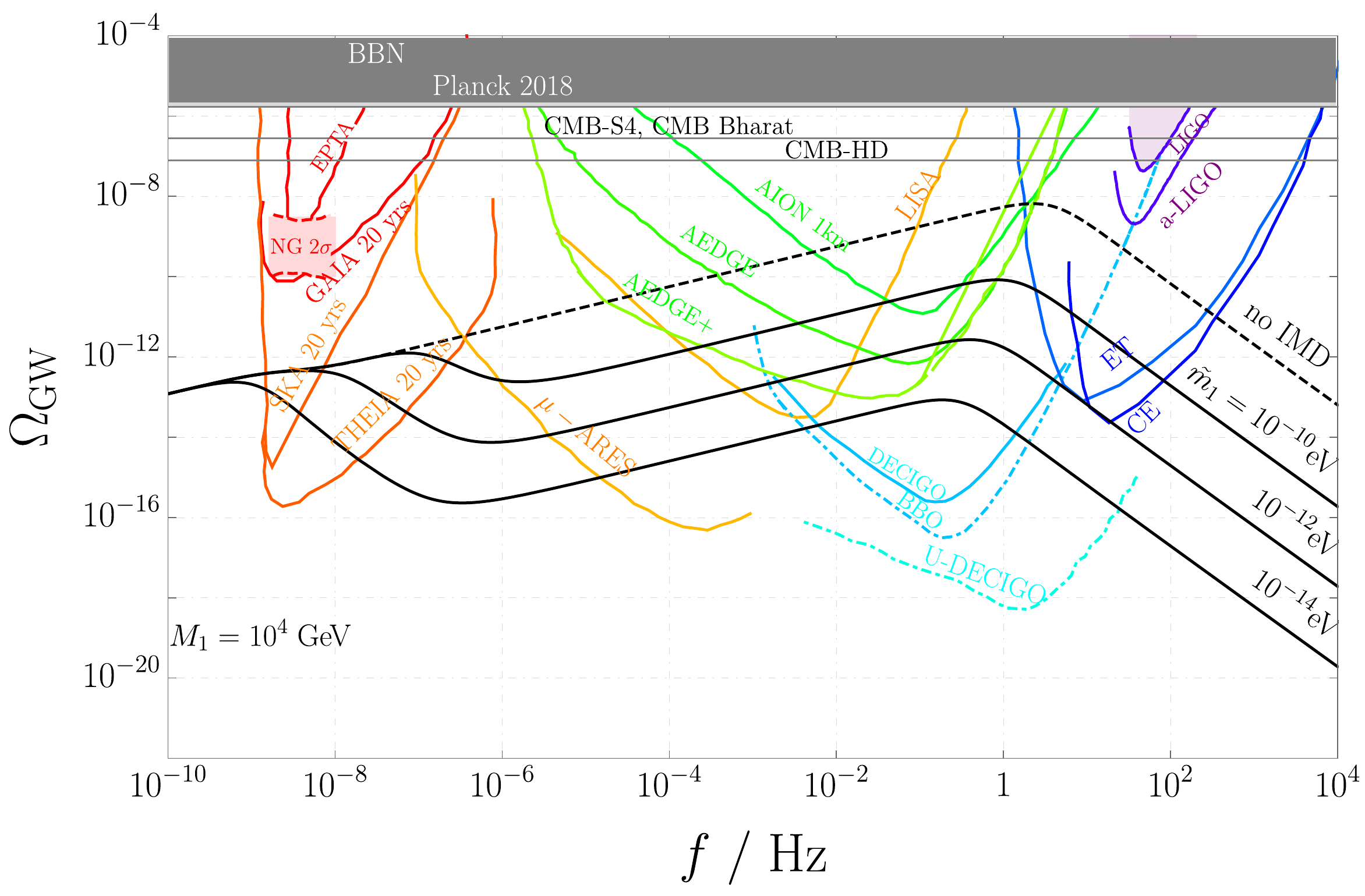}
    \caption{Example GW spectra for $T_\text{RH}=10^{8}\;\text{GeV}, \; M_1=10^4\;\text{GeV}$ and $n_T=0$ \textit{(left)} as well as $n_T=0.5$ \textit{(right)}. Here we varied $\tilde{m}_1 = \left(10^{-10},10^{-12} ,10^{-14} \right) \;\text{eV}$ and \enquote{no IMD}  refers to the scenario without RHN domination.}
    \label{fig:my_label1}
    \includegraphics[width=0.45\textwidth]{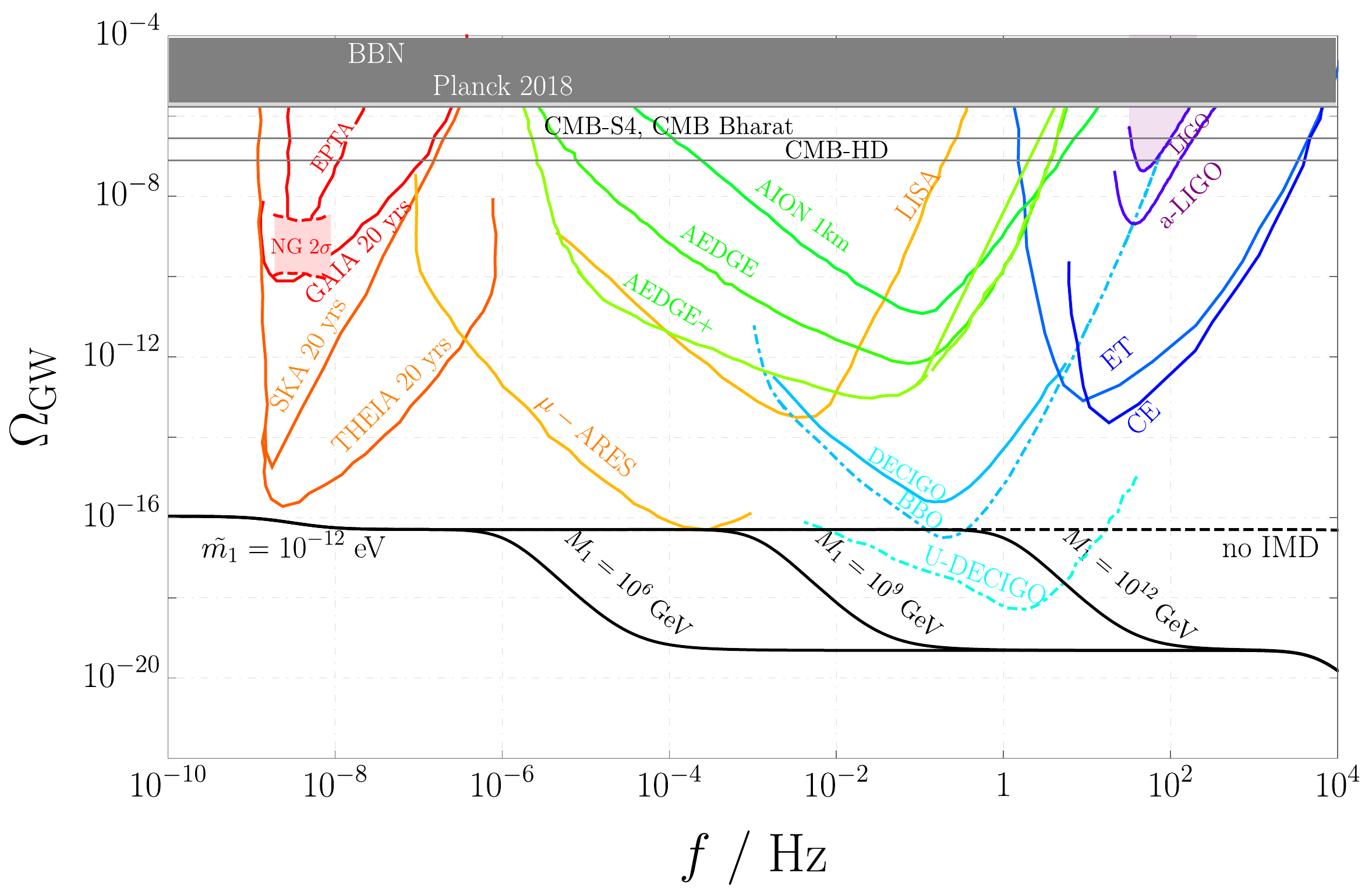}
    \includegraphics[width=0.45\textwidth]{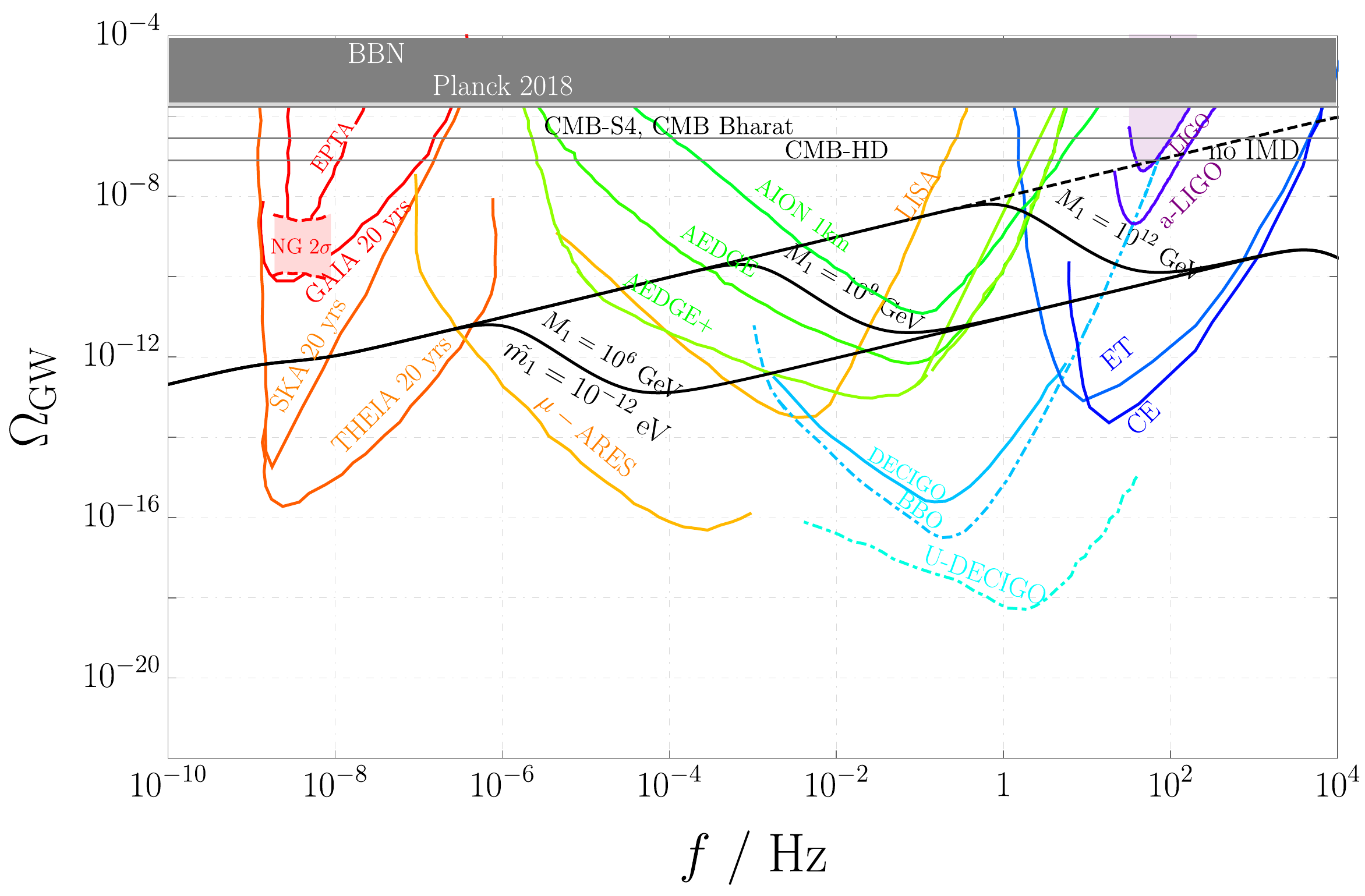}
    \caption{Example spectra for $T_\text{RH}=10^{12}\;\text{GeV}, \;\tilde{m}_1=10^{-12}\;\text{eV}$ and $n_T=0$ \textit{(left)} as well as $n_T=0.5$ \textit{(right)}. Here we varied $M_1=\left(10^6, 10^9, 10^{12}\right)\;\text{GeV}$ and \enquote{no IMD}  refers to the scenario without RHN domination.}
    \label{fig:my_label2}
\end{figure}

\subsection{Impact of free-streaming particles}
As shown in the seminal work \cite{Weinberg:2003ur} and expanded upon in \textit{e.g.} \cite{Watanabe:2006qe,Stefanek:2012hj,Dent:2013asa,Hook:2020phx}, there is a damping effect on the GW amplitude from free-streaming particles whose mean free path is larger than the Hubble scale. Free streaming particles such as the active neutrinos, the RHN, additional sources of dark radiation or gravitational waves themselves contribute to anisotropic stress-energy tensor and can reduce the primordial GW amplitude by up to 35.6\% \cite{Weinberg:2003ur}. In this work we neglect this effect to focus on the damping from the RHN induced matter dominated epoch as a first estimate, since percent level effects will only become relevant once we have actual data.

\section{Results}\label{sec:4}
\subsection{General results}\label{sec:res}

\begin{figure}
    \centering
    \includegraphics[width=0.95\textwidth]{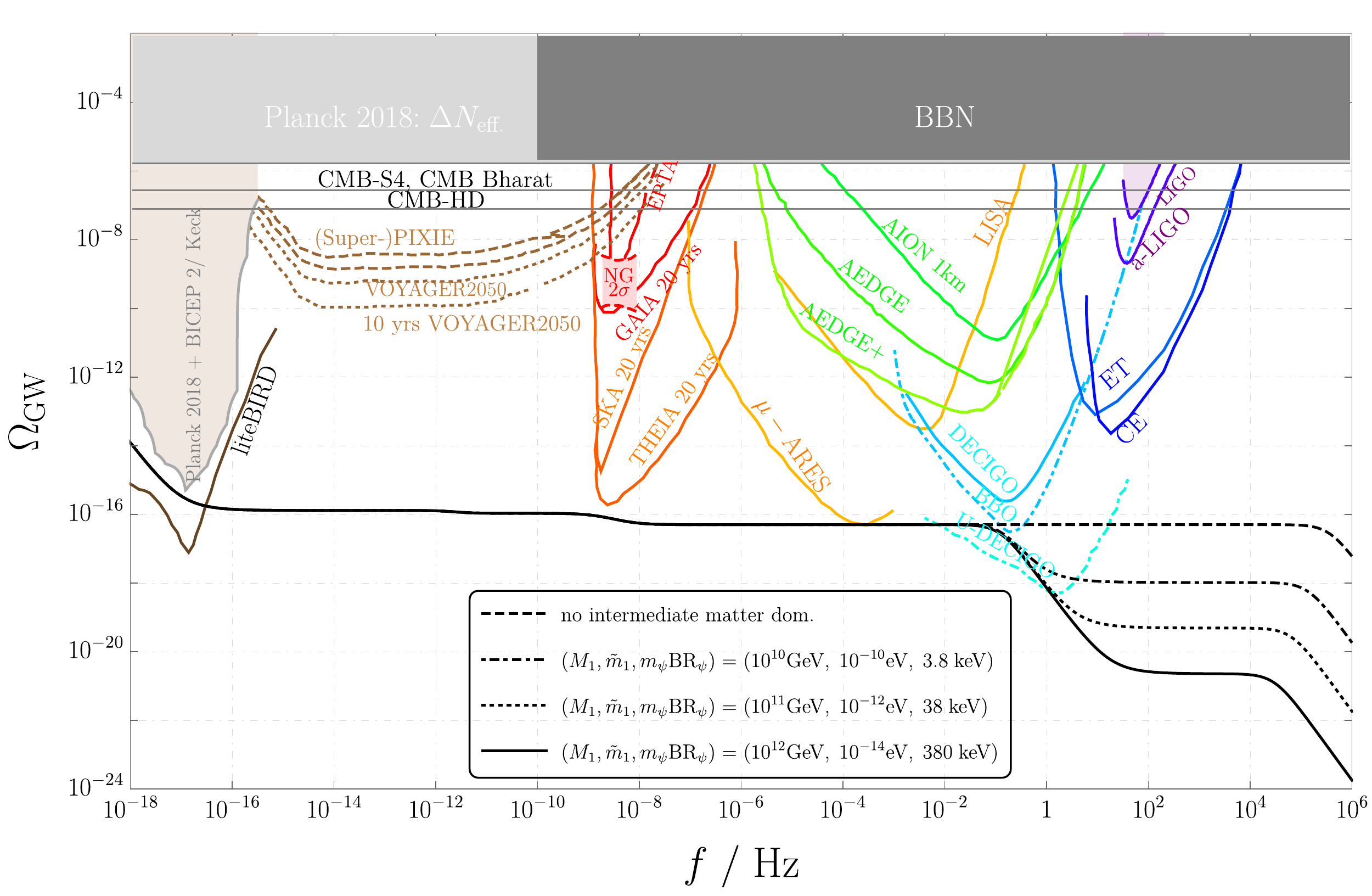}
     \caption{We fix $M_1$ as a function of $\tilde{m}_1=\left(10^{-10},10^{-12} ,10^{-14} \right) \;\text{eV}$ for successful leptogenesis and set $T_\text{RH}=10^{13}\;\text{GeV},\; n_T=0$. Furthermore we show which value of $m_\psi \text{BR}_\psi$ would be required for a given $\tilde{m}_1$ to generate the observed dark matter relic abundance.}
    \label{fig:my_label22}
    \includegraphics[width=0.95\textwidth]{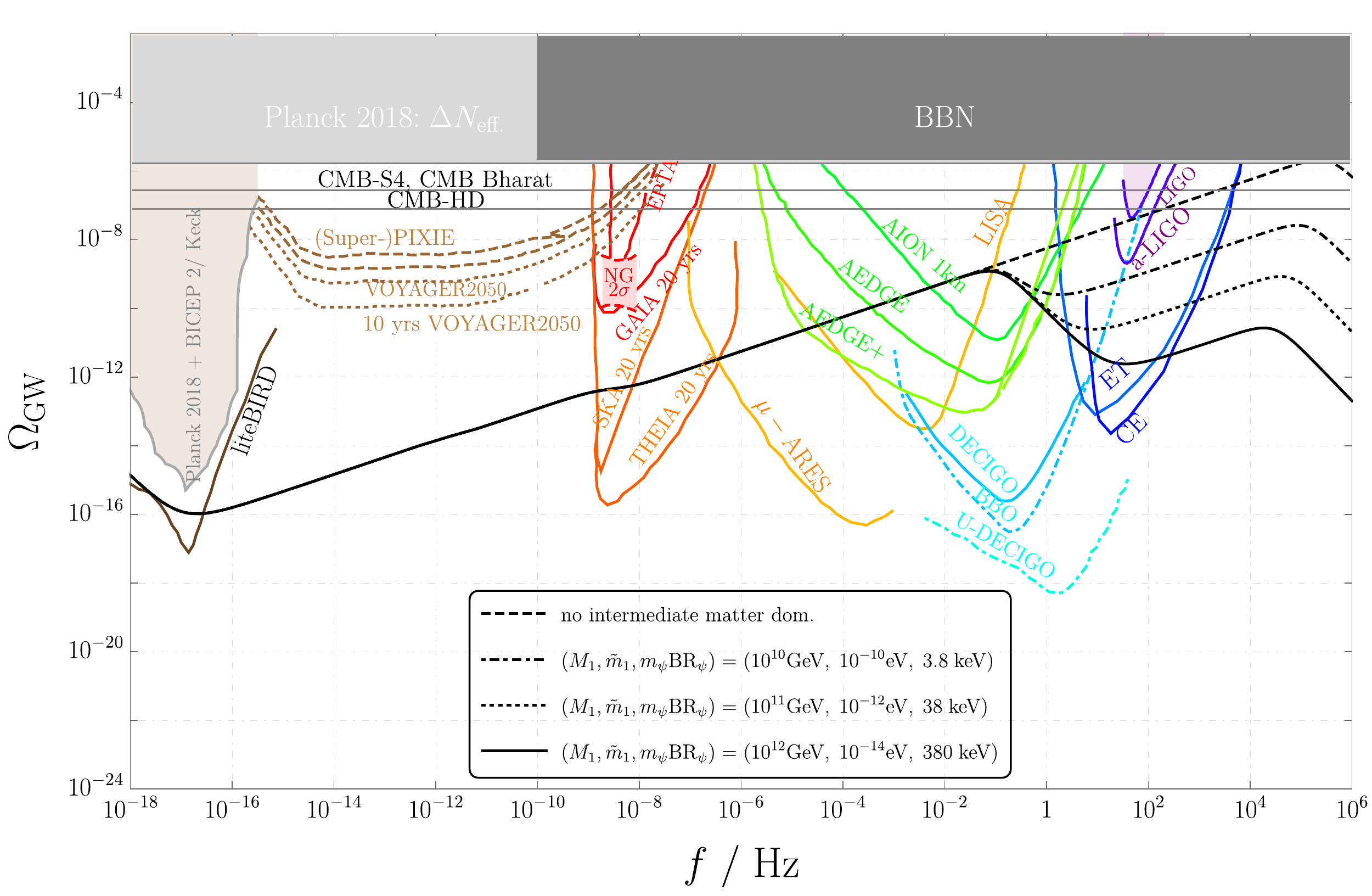}
    \caption{We fix $M_1$ as a function of $\tilde{m}_1=\left(10^{-10},10^{-12} ,10^{-14} \right) \;\text{eV}$ for successful leptogenesis and set $T_\text{RH}=10^{13}\;\text{GeV},\; n_T=0.5$. Furthermore we show which value of $m_\psi \text{BR}_\psi$ would be required for a given $\tilde{m}_1$ to generate the observed dark matter relic abundance.}
    \label{fig:my_label3}
\end{figure}

\begin{figure} 
    \centering
    \includegraphics[width=0.95\textwidth]{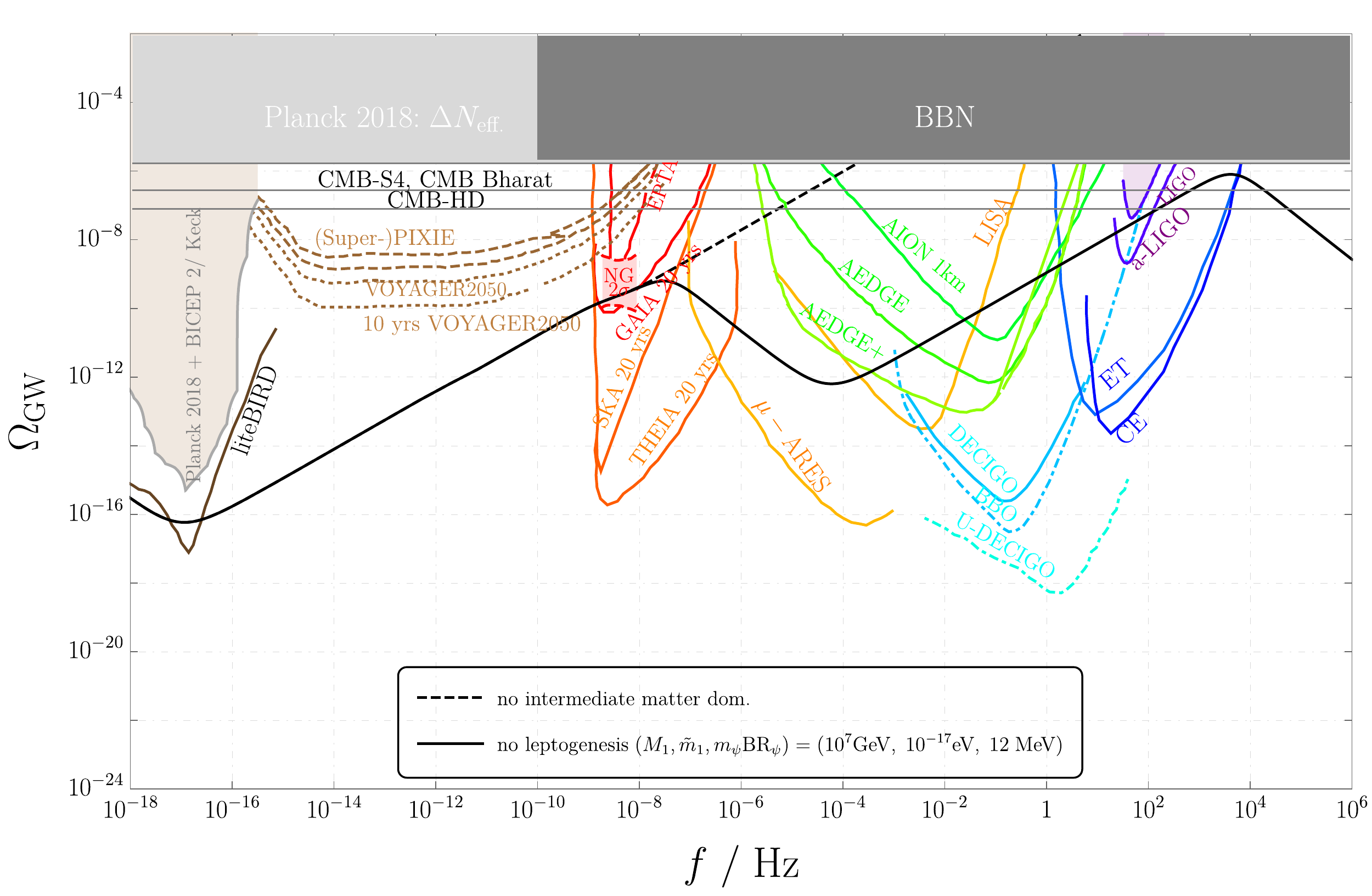}
    \caption{We fix $M_1=10^7\;\text{GeV}, \; \tilde{m}_1 = 10^{-17}\;\text{eV},\;T_\text{RH}=5\times10^{12}\;\text{GeV}$ and $n_T=0.85$ to fit the \textsc{NANOGRAV} anomaly \cite{NANOGrav:2020bcs}. Furthermore we show the  value of $m_\psi \text{BR}_\psi=\SI{12}{\mega\electronvolt}$  required for the given $\tilde{m}_1$ to generate the observed dark matter relic abundance.}
    \label{fig:nano}
\end{figure}

In the following we fix $r=0.035$ \cite{BICEP:2021xfz} and vary the reheating temperature as well as $M_1, \tilde{m}_1$ together with $n_T\geq 0$. We depict some example spectra in figures \ref{fig:my_label1} and \ref{fig:my_label2}, where we reproduced the figures from reference \cite{Asaka:2020wcr}. We depict the constraints from \textsc{LIGO}/\textsc{VIRGO}             \cite{LIGOScientific:2016aoc,LIGOScientific:2016sjg,LIGOScientific:2017bnn,LIGOScientific:2017vox,LIGOScientific:2017ycc,LIGOScientific:2017vwq} and \textsc{NANOGRAV} \cite{McLaughlin:2013ira,NANOGRAV:2018hou,Aggarwal:2018mgp,Brazier:2019mmu,NANOGrav:2020bcs} observations, the CMB as well as BBN as shaded regions in our plots \ref{fig:my_label1}-\ref{fig:nano}.
It is important to note that the depicted projection for the sensitivity of \textsc{U-DECIGO} \cite{Seto:2001qf,Kudoh:2005as,Kawamura_2006,Nakayama:2009ce,Yagi:2011wg,Kawamura:2020pcg}  is optimistic, but we do not employ the most optimistic  case known as \textsc{U-DECIGO-corr}, which assumes that the noise of the instrument is only given by the irreducible quantum noise \cite{Kudoh:2005as} and should therefore treated as a hypothetical best case scenario. The proposal for \textsc{BBO} \cite{Crowder:2005nr,Corbin:2005ny,Harry_2006} is also a bit speculative, because it is supposed to eventually succeed the currently planned \textsc{LISA} mission \cite{2017arXiv170200786A,Baker:2019nia}. To remind the reader of these potential caveats we depict the sensitivities for \textsc{U-DECIGO} and \textsc{BBO} with dashed-dotted lines in the figures \ref{fig:my_label1}-\ref{fig:nano}. The plots in figures \ref{fig:my_label22} and \ref{fig:my_label3} depict the case where we fix $M_1$ as function of $\tilde{m}_1$ according to \eqref{eq:M1} in order to reproduce the observed baryon asymmetry via leptogenesis.   In the aforementioned plot we also depict which values of  $m_\psi \text{BR}_\psi$ would be needed according to \eqref{eq:DM} to fit the dark matter relic abundance for a given $\tilde{m}_1$.
The labels \enquote{no IMD} in \ref{fig:my_label1}, \ref{fig:my_label2} and \enquote{no intermediate matter dom.} in \ref{fig:my_label22}-\ref{fig:my_label3} refer to the scenario without RHN domination computed from \eqref{eq:stand}, where the only dilution arises from inflationary reheating.
One can clearly see in \ref{fig:my_label22} and \ref{fig:my_label3} that the primordial tensor modes get diluted by the entropy released in the RHN decay for frequencies above $f_\text{sup.}\simeq 0.1\;\text{Hz}$, see \eqref{eq:freq}.
Furthermore one can observe in \ref{fig:my_label22}-\ref{fig:nano}
that there is  second break in the spectra at frequencies larger than $f_\text{sup.}\sim T_\text{dec.}$. This is due to the inflationary reheating at $T_\text{RH}$ and since our scenario is defined by the regime $T_\text{dec.} < M_1 < T_\text{RH}$ the second break occurs at a larger frequency. The same figures also show a small subleading suppression of frequencies larger than $\mathcal{O}(10^{-9}\;\text{Hz})$, which is due to the entropy released in the QCD phase transition \cite{Hajkarim:2019csy}.
Irrespective of the value of $n_T$, one can deduce from \ref{fig:my_label1}-\ref{fig:nano} that \textsc{LiteBIRD}~\cite{Hazumi:2019lys} will already probe the inflationary tensor modes in the $\left(10^{-16}-10^{-18}\right)\;\text{Hz}$ range. For $n_T=0$ we find that \textsc{U-DECIGO} \cite{Seto:2001qf,Kudoh:2005as,Kawamura_2006,Nakayama:2009ce,Yagi:2011wg,Kawamura:2020pcg} has the best chance to distinguish our entropy suppressed spectra from the standard case without RHN domination depicted by the dashed line in \ref{fig:my_label22}.
In case neither \textsc{BBO} \cite{Crowder:2005nr,Corbin:2005ny,Harry_2006} nor \textsc{U-DECIGO} \cite{Seto:2001qf,Kudoh:2005as,Kawamura_2006,Nakayama:2009ce,Yagi:2011wg,Kawamura:2020pcg} detect the tensor mode background expected from inflation, this does not have to rule out primordial gravitational waves and could be a tell-tale sign of scenarios with entropy dilution, such as ours. In the next section we will analyze this in terms of the SNR. The case of $n_T=0.5$ without RHN domination  would start to be probed by the dark radiation bounds in  \eqref{eq:darkrad2} from BBN  \cite{Cyburt:2015mya} and Planck  \cite{Planck:2018vyg} (see the dashed line in  \ref{fig:my_label3}) and is only borderline compatible with the existing \textsc{LIGO}/\textsc{VIRGO}         ~\cite{LIGOScientific:2016aoc,LIGOScientific:2016sjg,LIGOScientific:2017bnn,LIGOScientific:2017vox,LIGOScientific:2017ycc,LIGOScientific:2017vwq}  observations . An attempt to explain the recent anomaly  in the 12.5-year dataset \cite{NANOGrav:2020bcs} of the \textsc{NANOGRAV} collaboration  \cite{McLaughlin:2013ira,NANOGRAV:2018hou,Aggarwal:2018mgp,Brazier:2019mmu} with primordial tensor modes would require an extremely large $n_T\simeq 0.85$. The challenge is then to have enough entropy dilution to comply with the dark radiation and \textsc{LIGO}/\textsc{VIRGO}  bounds. We depict a spectrum for $M_1=10^7\;\text{GeV}, \; \tilde{m}_1 = 10^{-17}\;\text{eV}$ that could be the source of the anomaly in figure \ref{fig:nano} for the case without leptogenesis. The reason for abandoning leptogenesis  is simply that  with such a large $n_T$ the peak of the GW energy density at the typical frequency  $f_\text{sup.}=\SI{0.1}{\hertz}$ (before the dilution kicks in) will already be far too large to comply with the  dark radiation bounds. Therefore one needs a spectrum where the damping (which is only proportional to $\tilde{m}_1$ see \eqref{eq:dil}) occurs at lower decay temperatures and hence lower frequencies (set by both $M_1$ and $\tilde{m}_1$ see \eqref{eq:Tdec}). This is why we chose a value of $M_1=10^7\;\text{GeV}$ below the leptogenesis bound in \eqref{eq:M1}. On top of that we set $T_\text{RH}=\SI{5e12}{\giga\electronvolt}$, so that the GWs at large frequencies beyond \textsc{LIGO}/\textsc{VIRGO}          do not come into tension with the dark radiation bound due to the damping from inflationary reheating. These estimates illustrate, why we would need a rather contrived scenario and we do not pursue the aforementioned anomaly further in this work.

\subsection{Signal-to-noise ratio}\label{sec:SNR}

\begin{figure}[t]
    \centering
    \includegraphics[width=0.45\textwidth]{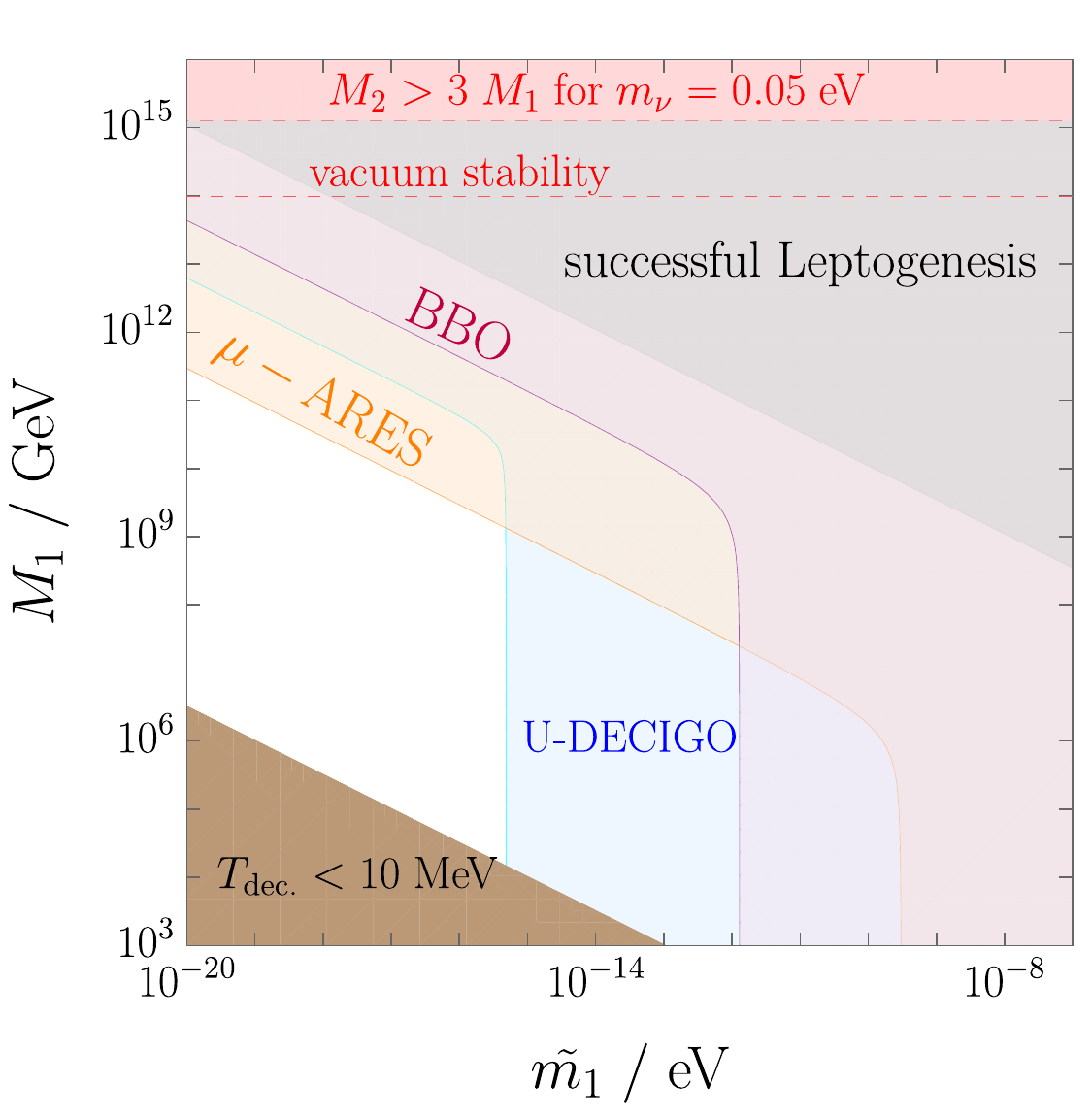}
    \includegraphics[width=0.5\textwidth]{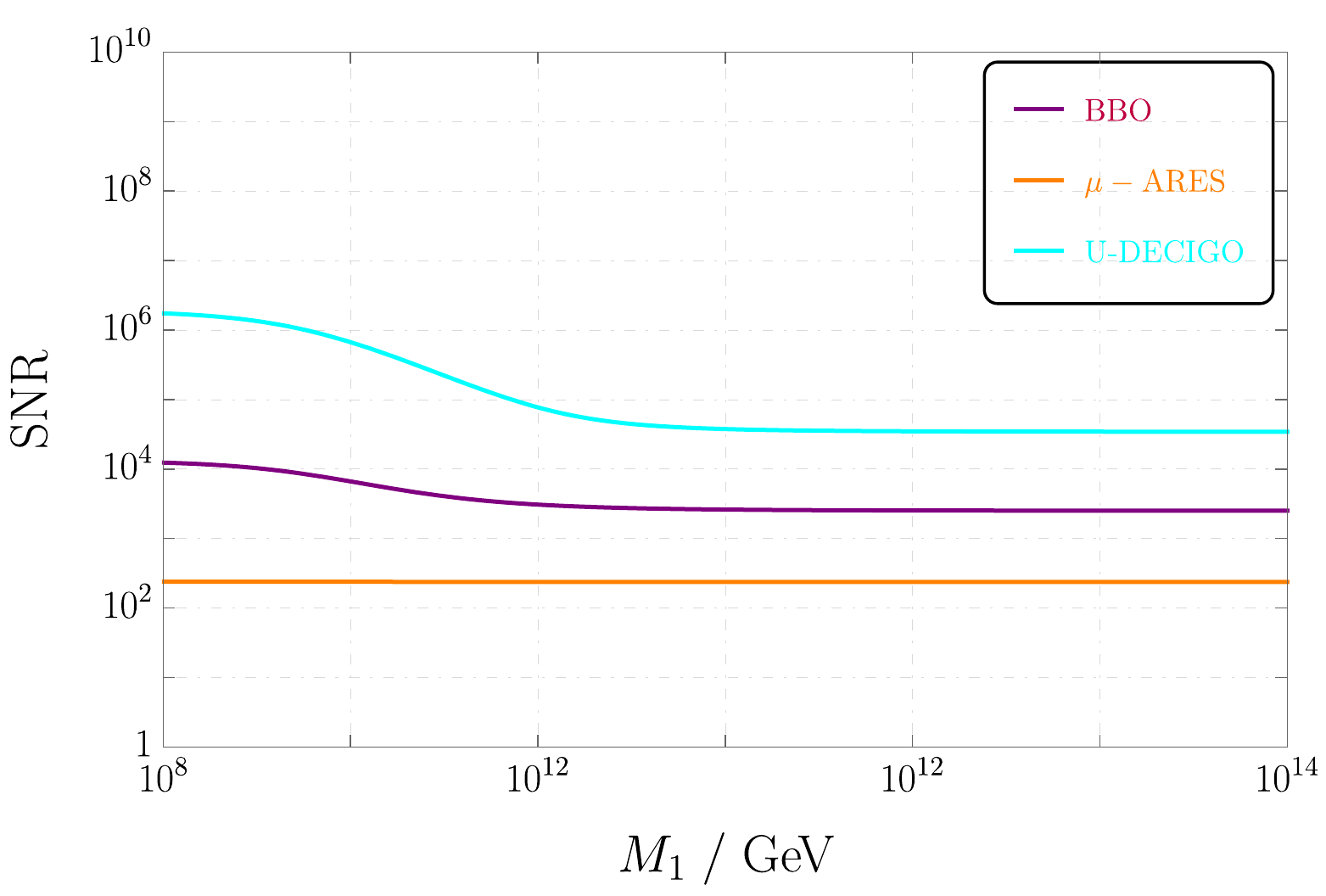}
    \caption{Parameter space in the $M_1$ versus $\tilde{m}_1$ plane with contours for $\text{SNR} =10$ \textit{(left)} and SNR as a function of $M_1$, where $\tilde{m}_1$ was fixed for leptogenesis via  \eqref{eq:M1} \textit{(right)}. In both plots we fixed $T_\text{RH}=10^{16}\;\text{GeV},\; n_T=0$.
    See the main text for details on the constraints.
    The SNR is larger than 10 in the colored regions.
    Note that the colored lines from the experiments do no correspond to constraints, but to projections of future sensitivities.
    }
    \label{fig:my_label4}
\end{figure}

\begin{figure}[t]
    \centering
    \includegraphics[width=0.45 \textwidth]{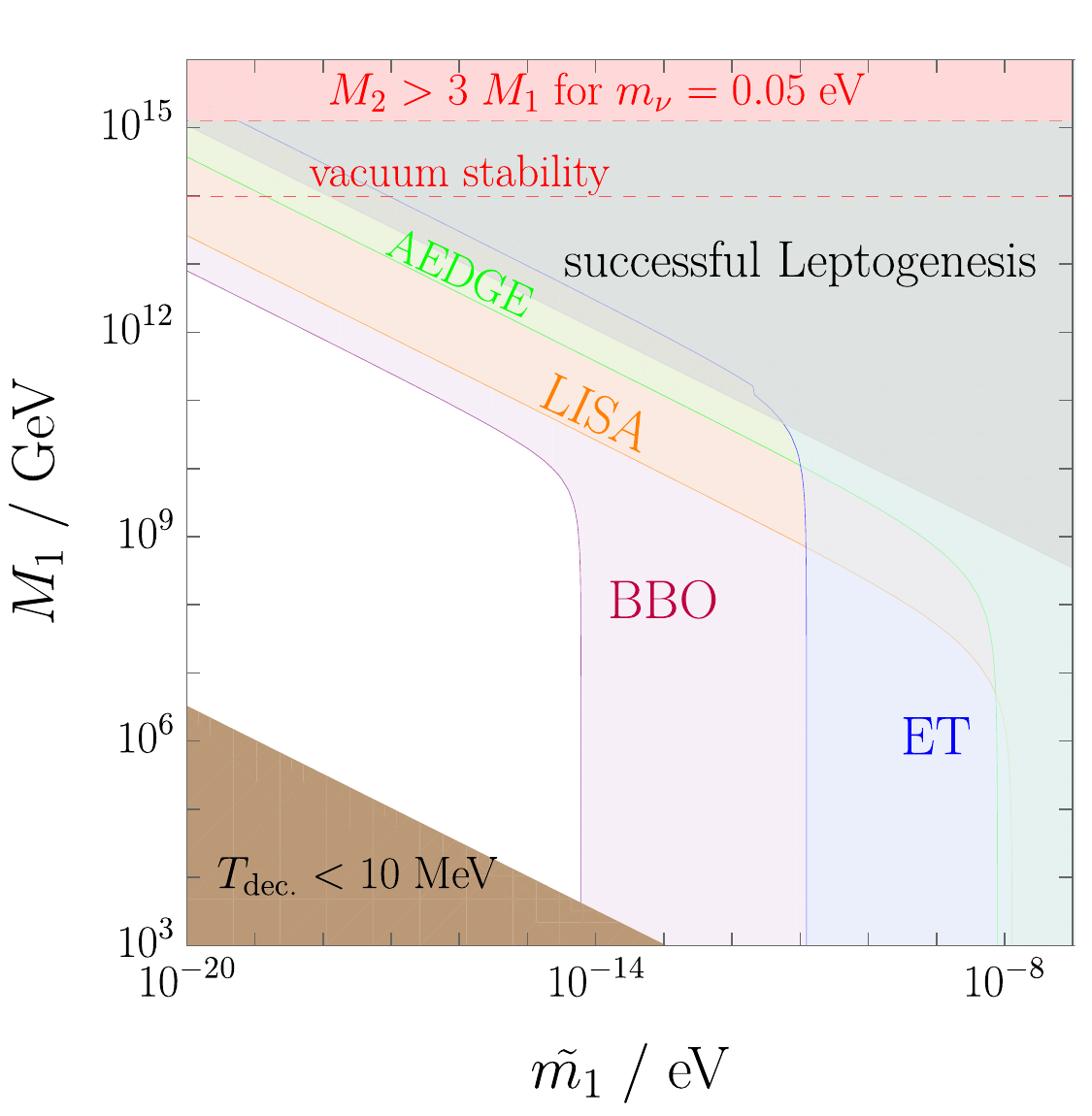}
    \includegraphics[width=0.45 \textwidth]{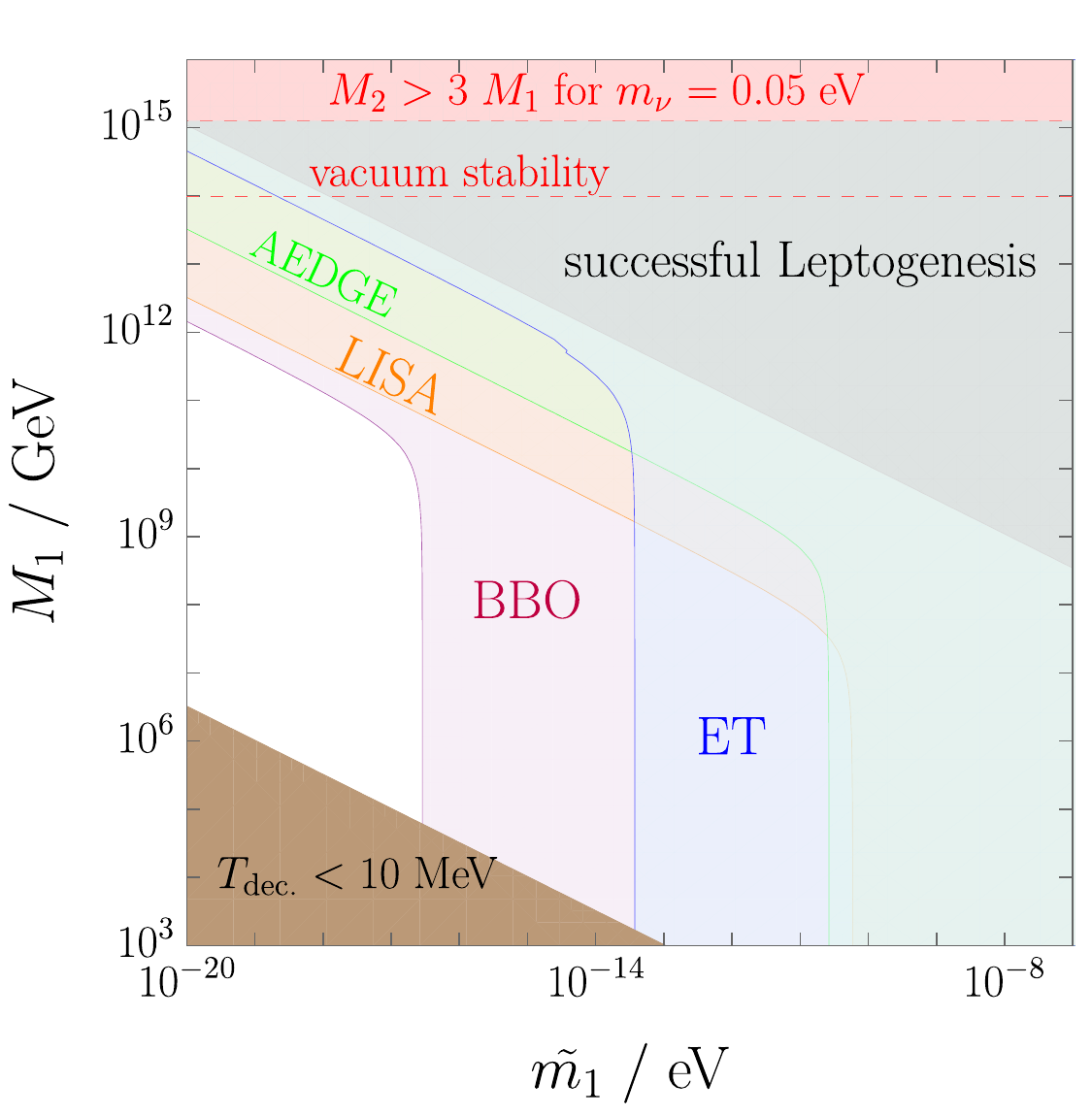}
    \caption{Parameter space in the $M_1$ versus $\tilde{m}_1$ plane with contours for $\text{SNR} =10$
    for $n_T=0.1$  \textit{(left)}  and  $n_T=0.2$ \textit{(right)}. In both plots we fixed $T_\text{RH}=10^{16}\;\text{GeV}$. See the main text for details on the constraints.
    The SNR is larger than 10 in the colored regions.
    Note that the colored lines from the experiments do no correspond to constraints, but to projections of future sensitivities.}
    \label{fig:my_label6}
    \includegraphics[width=0.45 \textwidth]{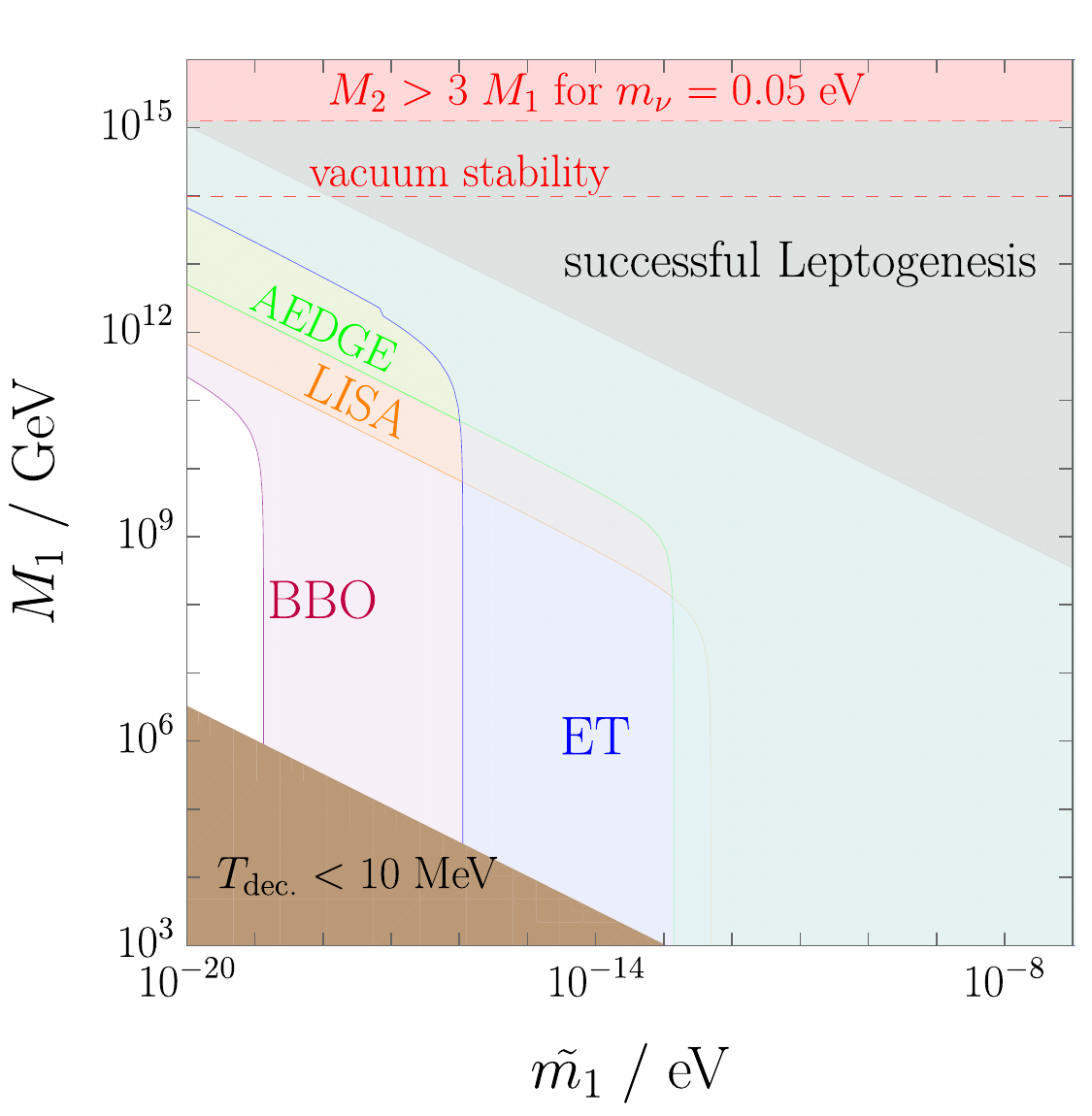}
    \includegraphics[width=0.45\textwidth]{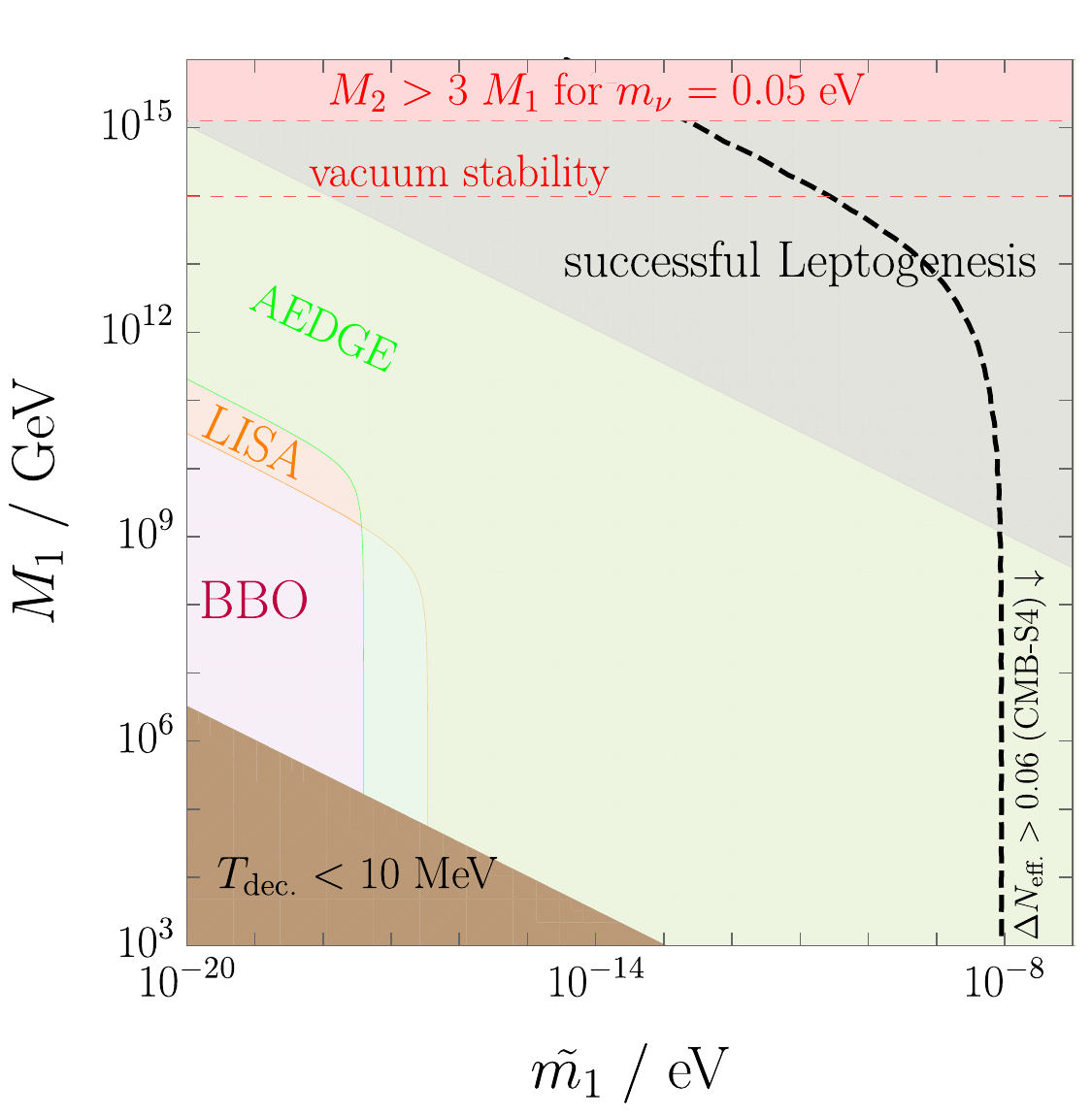}
    \caption{Parameter space in the $M_1$ versus $\tilde{m}_1$ plane with contours for $\text{SNR} =10$ 
    for $n_T=0.3$  \textit{(left)}  and  $n_T=0.5$ \textit{(right)}. In both plots we fixed $T_\text{RH}=10^{16}\;\text{GeV}$. See the main text for details on the constraints.
    The SNR is larger than 10 in the colored regions.
    Note that the colored lines from the experiments do no correspond to constraints, but to projections of future sensitivities.}
    \label{fig:my_label8}
\end{figure}

\begin{figure}[t]
    \centering
    \includegraphics[width=0.45\textwidth]{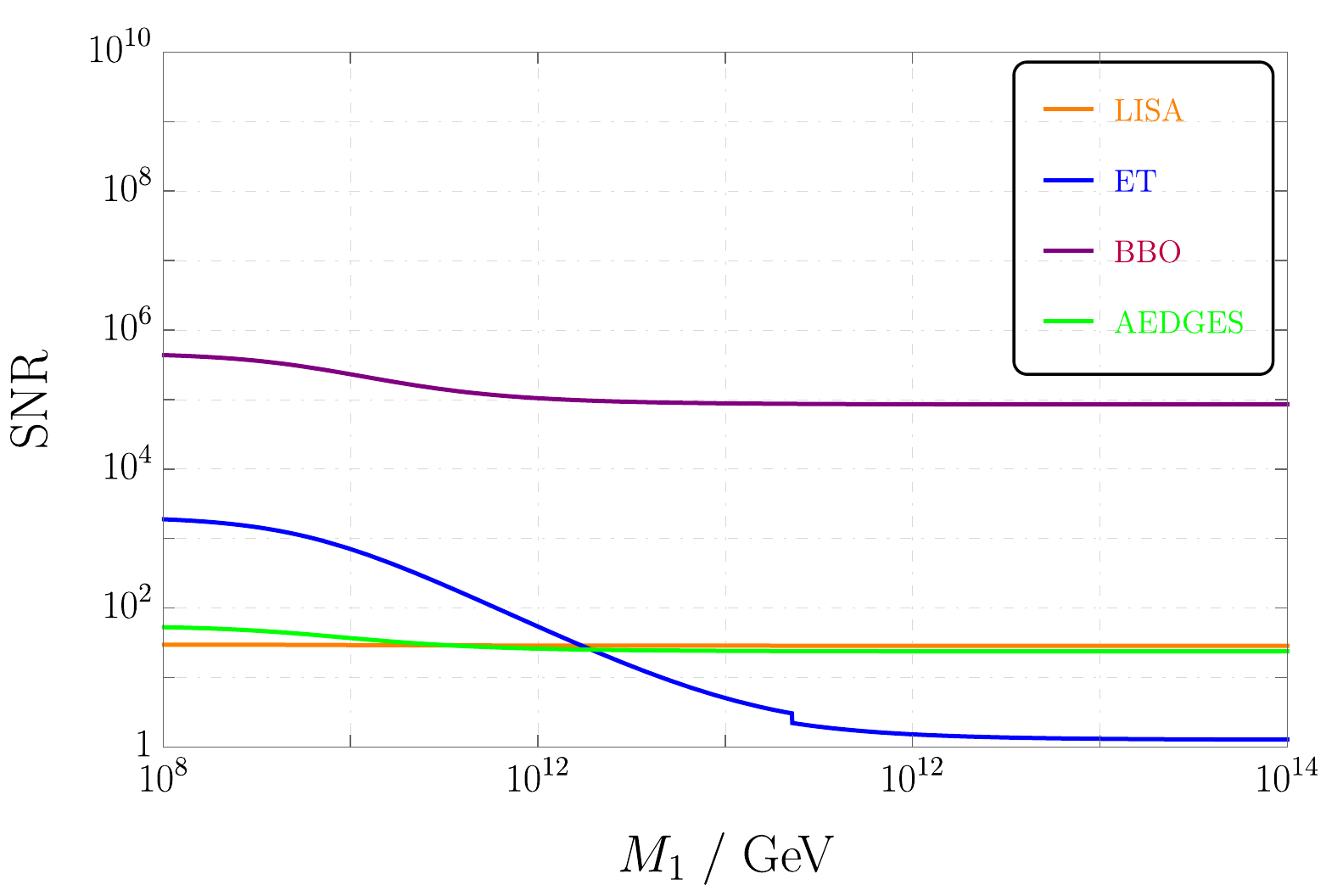}
    \includegraphics[width=0.45\textwidth]{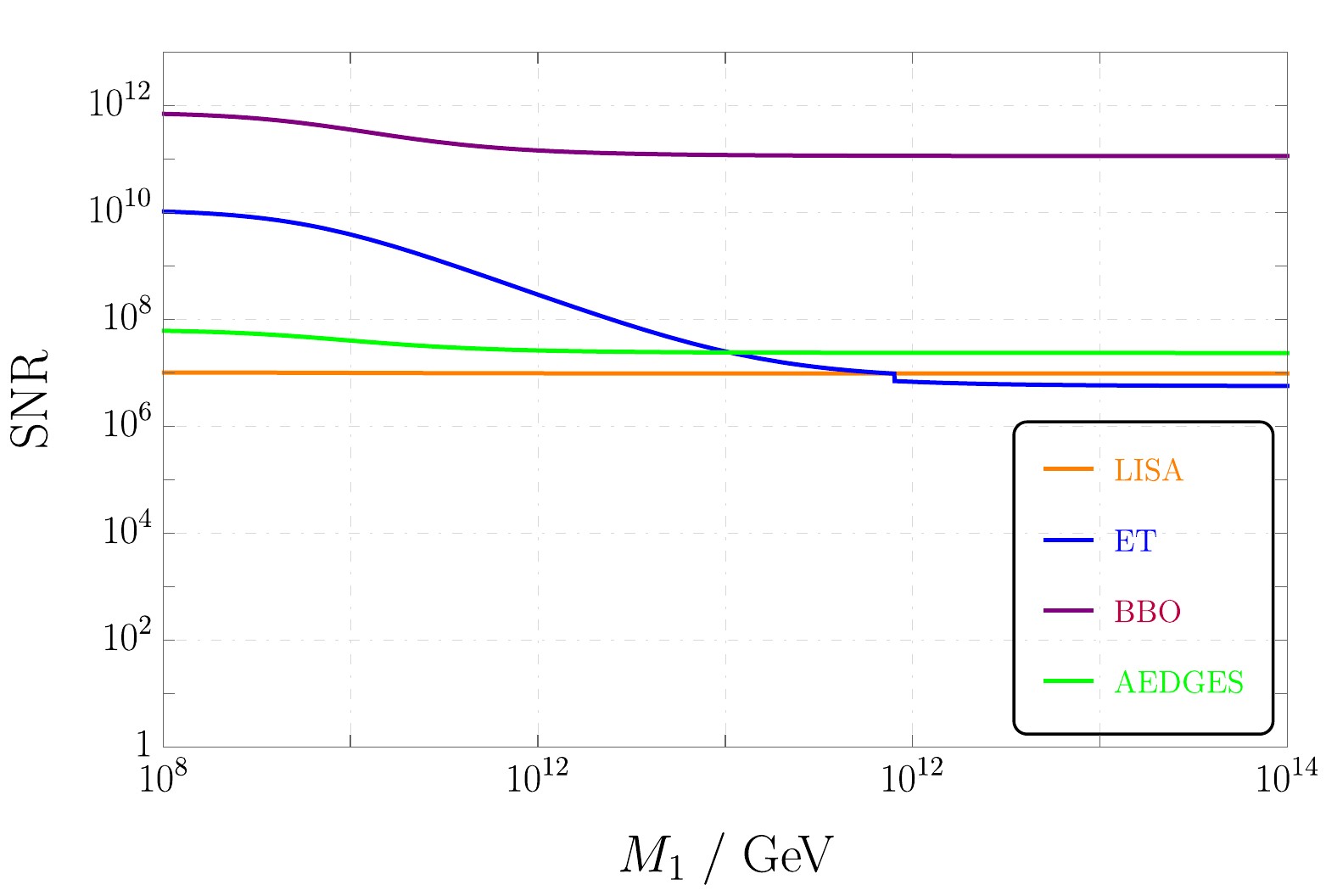}
    \caption{SNR as a function of $M_1$, where $\tilde{m}_1$ was fixed for leptogenesis via  \eqref{eq:M1} with $n_T=0.1$ \textit{(left)} and $n_T=0.5$ \textit{(right)}. In both plots we fixed $T_\text{RH}=10^{16}\;\text{GeV}$ }
    \label{fig:my_label11}
\end{figure}

We use the SNR defined in \eqref{eq:SNR} to determine the region in the $M_1$ versus $\tilde{m}_1$ parameter space, where a detection of primordial gravitational waves can be claimed for a SNR threshold of ten over four years of observation time. For $n_T=0$ we find that \textsc{BBO} \cite{Crowder:2005nr,Corbin:2005ny,Harry_2006},  \textsc{$\mu$-ARES} \cite{Sesana:2019vho} and \textsc{U-DECIGO} \cite{Seto:2001qf,Kudoh:2005as,Kawamura_2006,Nakayama:2009ce,Yagi:2011wg,Kawamura:2020pcg} are the most relevant experiments that have a chance of probing the primordial GW background, as can be deduced from figure  \ref{fig:my_label22}.
For $n_T>0$ there are a lot more experiments that can probe our GW spectra, which is why we focus on \textsc{AEDGE} \cite{AEDGE:2019nxb,Badurina:2021rgt}, \textsc{BBO} \cite{Crowder:2005nr,Corbin:2005ny,Harry_2006}, the \textsc{Einstein Telescope (ET)} \cite{Punturo:2010zz,Hild:2010id} and \textsc{LISA} \cite{2017arXiv170200786A,Baker:2019nia}. Of course there are also other currently developed experiments, such as the radio telescope \textsc{SKA}~\cite{Carilli:2004nx,Janssen:2014dka,Weltman:2018zrl}, that become relevant for $n_T>0$.
The parameter space for $n_T=0$ was displayed in \ref{fig:my_label4}, whereas figure \ref{fig:my_label6} showcases $n_T=0.1, \;0.2$ and \ref{fig:my_label8} the cases of 
$n_T=0.3, \;0.5$. The region in \eqref{eq:M1} that leads to the observed baryon asymmetry via leptogenesis was shaded in gray.  For $n_T=0.1$ one can conclude from the left plot in figure \ref{fig:my_label6} that the SNR threshold for \textsc{ET} \cite{Punturo:2010zz,Hild:2010id} will start to probe the edge of the parameter space for leptogenesis in the regime $\tilde{m}_1\lesssim 10^{-11}\;\text{eV}$. For $n_T>0$ we see in \ref{fig:my_label6}-\ref{fig:my_label8} that  \textsc{AEDGE}~\cite{AEDGE:2019nxb,Badurina:2021rgt}, \textsc{BBO}~ \cite{Crowder:2005nr,Corbin:2005ny,Harry_2006} and \textsc{LISA} ~\cite{2017arXiv170200786A,Baker:2019nia}   probe the entire parameter space for leptogenesis. We impose the following constraints in figures  \ref{fig:my_label4}-\ref{fig:my_label8}: 
Successful BBN requires that the RHN decay temperature in \eqref{eq:Tdec} is at least $10\;\text{MeV}$ \cite{Kawasaki:2000en,Hannestad:2004px}, which was depicted as a brown region.
RHN with masses above $10^{14}\;\text{GeV}$ could destabilize the electroweak vacuum \cite{Casas:1999cd,Elias-Miro:2011sqh}. 
We do not show the bound $M_1\lesssim 10^7\;\text{GeV}$ \cite{Vissani:1997ys,Clarke:2015gwa,Brivio:2017dfq,Brivio:2018rzm}  from the naturalness of the Higgs mass under corrections from its couplings to the RHN, as it would basically exclude   our entire parameter space in \eqref{eq:M1}.
The last bound comes from  the observed neutrino masses: Due to the perturbativity of the RHN Yukawa coupling $\lambda_{ij}<\sqrt{4\pi}$ and the need to reproduce at least one mass eigenstate with $m_\nu=\SI{0.05}{\electronvolt}$ we find that $M_2 \lesssim \SI{3.8e15}{\giga\electronvolt}$. This together with our assumption that $M_{2} > 3 M_1$ means that we have to require at least $M_1 \lesssim 10^{15}\;\text{GeV}$. In all plots we fixed $T_\text{RH} = 10^{16}\;\text{GeV}$ so that even the heaviest  $N_1$ allowed by the previous considerations would be present in the plasma. As mentioned in the previous section we find that \textsc{U-DECIGO} \cite{Seto:2001qf,Kudoh:2005as,Kawamura_2006,Nakayama:2009ce,Yagi:2011wg,Kawamura:2020pcg} is the best candidate to test our setup compared to the case with no decaying RHN for $n_T=0$. A future non-observation of the inflationary tensor mode spectrum could be explained by a decaying $N_1$ with $\tilde{m}_1 <10^{-14}\;\text{eV}$ and a mass of $M_1\gtrsim 10^4\;\text{GeV}$ (the precise number depends on the BBN bound on the RHN decay temperature of at least $10\;\text{MeV}$). By fixing $\tilde{m}_1$ as a function of $M_1$ for leptogenesis via  \eqref{eq:M1} we plot the SNR as a function of $M_1$ on the right side of \ref{fig:my_label4}. 
Here the SNR for  \textsc{$\mu$-ARES} \cite{Sesana:2019vho} is constant because the peak of its sensitivity is situated at a frequency below $f_\text{sup.}\simeq0.1\;\text{Hz}$ and it is therefore blind to the entropy damping.  
For cosmologies with $n_T>0$ we find that the SNR for \textsc{U-DECIGO} \cite{Seto:2001qf,Kudoh:2005as,Kawamura_2006,Nakayama:2009ce,Yagi:2011wg,Kawamura:2020pcg} is always larger than 10 in the depicted parameter space, which is why we focus on different detectors.
\textsc{BBO} \cite{Crowder:2005nr,Corbin:2005ny,Harry_2006} is a  promising candidate for a detection of primordial GWs with both $n_T=0$ and $n_T>0$ (compare the plots in \ref{fig:my_label4} and \ref{fig:my_label6}, \ref{fig:my_label8}). For   $n_T\gtrsim 0.5$   the dark radiation bound becomes important again and we show the contour  $\Delta N_\text{eff}^\text{proj.}=0.06$ for    CMB Stage IV \cite{Abazajian:2019eic,annurev-nucl-102014-021908} computed via \eqref{eq:darkrad} on the right side of figure \ref{fig:my_label8}. For completeness we display the   SNR as a function of $M_1$ (with $\tilde{m}_1$  fixed by leptogenesis   \eqref{eq:M1}) for $n_T=0.1,\; 0.5$ in figure \ref{fig:my_label11}.

\section{Conclusions and Discussions}\label{sec:5}
We focused on the minimal Seesaw model, which adds only three right handed neutrinos (RHN) to the SM, and demonstrated that an epoch of  right handed neutrino domination, with a  Yukawa coupling corresponding to $\tilde{m}_1 < \SI{2.9e-7}{\electronvolt}$, can realize baryogenesis via leptogenesis for  a mass of $M_1 \gtrsim \SI{2.4e8}{\giga\electronvolt} \cdot \sqrt{\SI{2e-7}{\electronvolt}/\tilde{m}_1}$ (see \eqref{eq:M1}). Since the effective mass is $\tilde{m}_1$ is too small for a thermal RHN population, we had to assume a different production channel via either inflaton decays or B-L gauge scatterings for the initial RHN abundance. Furthermore such a small $\tilde{m}_1$ requires that one of the SM neutrinos is approximately massless compared to the other two. The amplitude of gravitational waves that re-enter the horizon before the end of the RHN matter dominated epoch is damped by a factor proportional to the entropy released in the RHN decay.  We discussed the detection possibilities of primordial GWs and computed the signal-to-noise ratio for various detectors  such as \textsc{AEDGE}  \cite{AEDGE:2019nxb,Badurina:2021rgt}, \textsc{BBO} \cite{Crowder:2005nr,Corbin:2005ny,Harry_2006}, \textsc{DECIGO}  \cite{Seto:2001qf,Kudoh:2005as,Kawamura_2006,Nakayama:2009ce,Yagi:2011wg,Kawamura:2020pcg}, \textsc{Einstein Telescope}  \cite{Punturo:2010zz,Hild:2010id}, \textsc{LISA} \cite{2017arXiv170200786A,Baker:2019nia} or \textsc{$\mu$-ARES} \cite{Sesana:2019vho}
as well as for several spectral tilts $n_T\geq 0$ of the tensor mode spectrum. Additionally we determined the regions in the $M_1$ versus $\tilde{m}_1$ parameter space in which the signal-to-noise ratio (SNR) is larger than ten over a four year observational period in the figures \ref{fig:my_label4}-\ref{fig:my_label8}. Our main finding is that high scale leptogenesis can have an observable imprint on the gravitational waves from inflation.  Further we discussed under which conditions our scenario leads to  the dominant GW signal.  Since fixing $M_1$ as a function of $\tilde{m}_1$ for successful leptogenesis by saturating the maximum of the CP-violating decay parameter $\varepsilon_1$ for a hierarchical spectrum (see \eqref{eq:max}) completely determines  the RHN decay temperature to be $T_\text{dec.} \simeq \SI{3.3e6}{\giga\electronvolt}$, we find a constant characteristic frequency of $f_\text{sup.}\simeq \SI{0.1}{\hertz}$ (see \eqref{eq:freq} and figures \ref{fig:my_label22}-\ref{fig:my_label3}), above which the suppression of the GW amplitude manifests itself.
The same RHN can also have a second potentially suppressed decay mode to a stable fermion $\psi$, that is responsible for the dark matter abundance, if the product of the DM mass and the branching fraction of the RHN decay to DM satisfies $m_\psi  \text{BR}_\psi \simeq \SI{85}{\electronvolt}\cdot  \sqrt{\SI{2e-7}{\electronvolt}/\tilde{m}_1}$. In order for the dark matter do be heavy enough for successful structure formation ($m_\psi >\mathcal{O}(\SI{10}{\kilo\electronvolt})$) for fixed $\tilde{m}_1$ we typically need a small branching ratio   $\text{BR}_\psi\ll 1$. Such a small branching fraction can also suppress the amount of BSM dark radiation $\Delta N_\text{eff.} \simeq 0.06 \cdot (\text{BR}_\psi / 4\%)$, that could potentially  be generated, if the scalar produced together with $\psi$ is very light and survives until today. This particular scenario  leads to GeV-scale DM decaying to dark radiation and SM neutrinos, which necessitates $ \tilde{m}_1 < \SI{9.7e-15}{\electronvolt}$ in order to have  DM with a large enough lifetime on cosmological  scales and the right relic abundance. 

\section*{Acknowledgements}
We would like to thank Bowen Fu,  Stephen King, Alessandro Strumia  and Andreas Trautner for  useful comments on the manuscript.

\bibliographystyle{JHEP}
\bibliography{ref}

\end{document}